\documentclass[acmtog, nonacm, screen]{acmart}

\usepackage{xcolor}
\usepackage{subcaption}
\usepackage{tikz}
\usepackage{multirow}
\usepackage{multicol}
\usepackage{tabularx}
\usepackage{booktabs}
\usepackage{nicefrac}
\usepackage[inline]{enumitem}
\usepackage{makecell}
\usepackage{microtype}
\usetikzlibrary{positioning}

\usepackage[capitalize]{cleveref}
\crefname{section}{sec.}{secs.}
\Crefname{section}{Sec.}{Secs.}
\crefname{table}{tab.}{tabs.}
\Crefname{table}{Tab.}{Tabs.}
\crefname{figure}{fig.}{figs.}
\Crefname{figure}{Fig.}{Figs.}
\crefname{equation}{eq.}{eqs.}
\Crefname{equation}{Eq.}{Eqs.}

\makeatletter
\def\thickhline{%
  \noalign{\ifnum0=`}\fi\hrule \@height \thickarrayrulewidth \futurelet
   \reserved@a\@xthickhline}
\def\@xthickhline{\ifx\reserved@a\thickhline
               \vskip\doublerulesep
               \vskip-\thickarrayrulewidth
             \fi
      \ifnum0=`{\fi}}
\makeatother

\newlength{\thickarrayrulewidth}
\begin{document}
\title{LM-GAN: A Photorealistic All-Weather Parametric Sky Model}

\author{Lucas Valença}
\affiliation{%
  \institution{Université Laval}
  \country{Canada}}
\email{lucas.valenca@ulaval.ca}

\author{Ian Maquignaz}
\affiliation{%
  \institution{Université Laval}
  \country{Canada}}
\email{ian.maquignaz.1@ulaval.ca}

\author{Hadi Moazen}
\affiliation{%
  \institution{Université Laval}
  \country{Canada}}
\email{hadi.moazen.1@ulaval.ca}

\author{Rishikesh Madan}
\affiliation{%
  \institution{Université Laval}
  \country{Canada}}
\email{rishikesh.madan.1@ulaval.ca}

\author{Yannick Hold-Geoffroy}
\affiliation{%
  \institution{Adobe Research}
  \country{USA}}
\email{holdgeof@adobe.com}

\author{Jean-François Lalonde}
\affiliation{%
  \institution{Université Laval}
  \country{Canada}}
\email{jean-francois.lalonde@gel.ulaval.ca}

\begin{abstract}
We present LM-GAN, an HDR sky model that generates photorealistic environment maps with weathered skies. 
Our sky model retains the flexibility of traditional parametric models and enables the reproduction of photorealistic all-weather skies with visual diversity in cloud formations. This is achieved with flexible and intuitive user controls for parameters, including sun position, sky color, and atmospheric turbidity.
Our method is trained directly from inputs fitted to real HDR skies, learning both to preserve the input's illumination and correlate it to the real reference's atmospheric components in an end-to-end manner. Our main contributions are a generative model trained on both sky appearance and scene rendering losses, as well as a novel sky-parameter fitting algorithm. We demonstrate that our fitting algorithm surpasses existing approaches in both accuracy and sky fidelity, and also provide quantitative and qualitative analyses, demonstrating LM-GAN's ability to match parametric input to photorealistic all-weather skies. The generated HDR environment maps are ready to use in 3D rendering engines and can be applied to a wide range of image-based lighting applications.
\end{abstract}

\begin{teaserfigure}
    \hbox{\put(0, 10){\includegraphics[width=.495\textwidth]{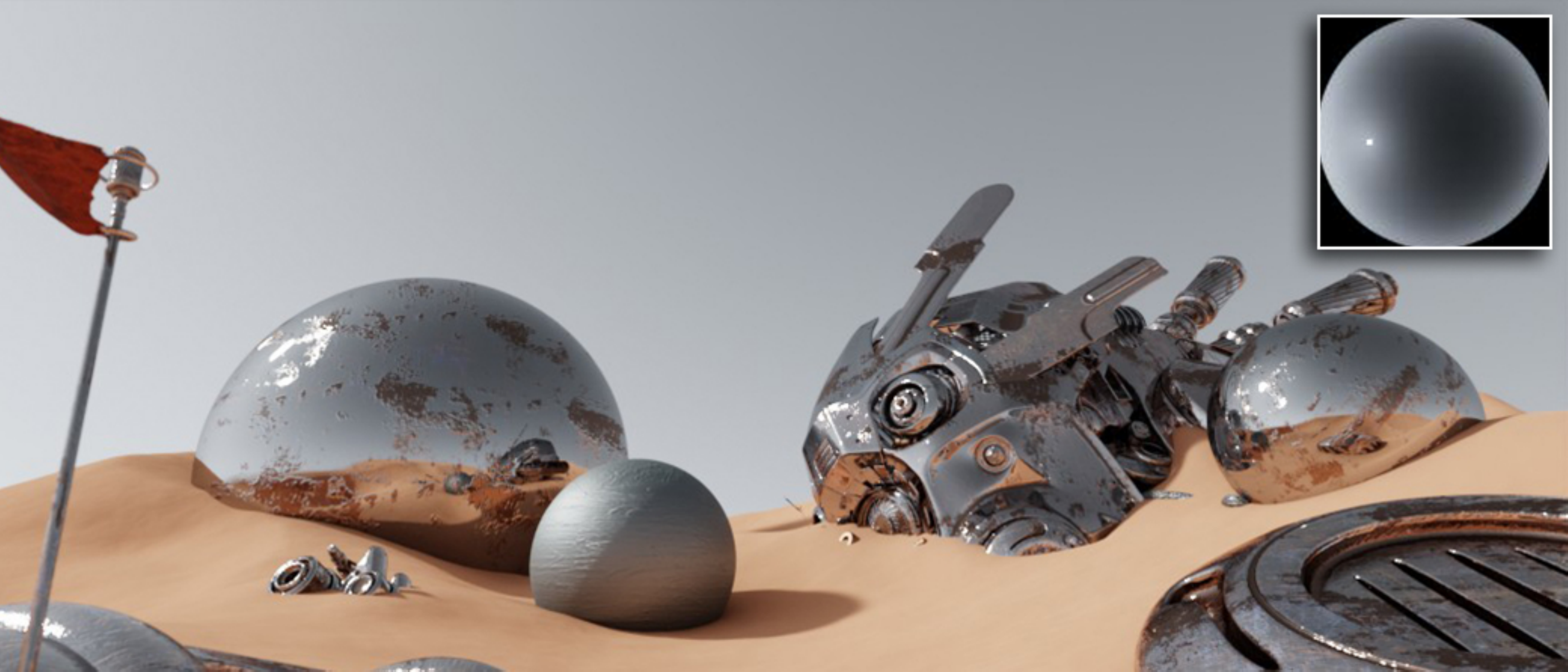}}
          \put(257, 10){\includegraphics[width=.495\textwidth]{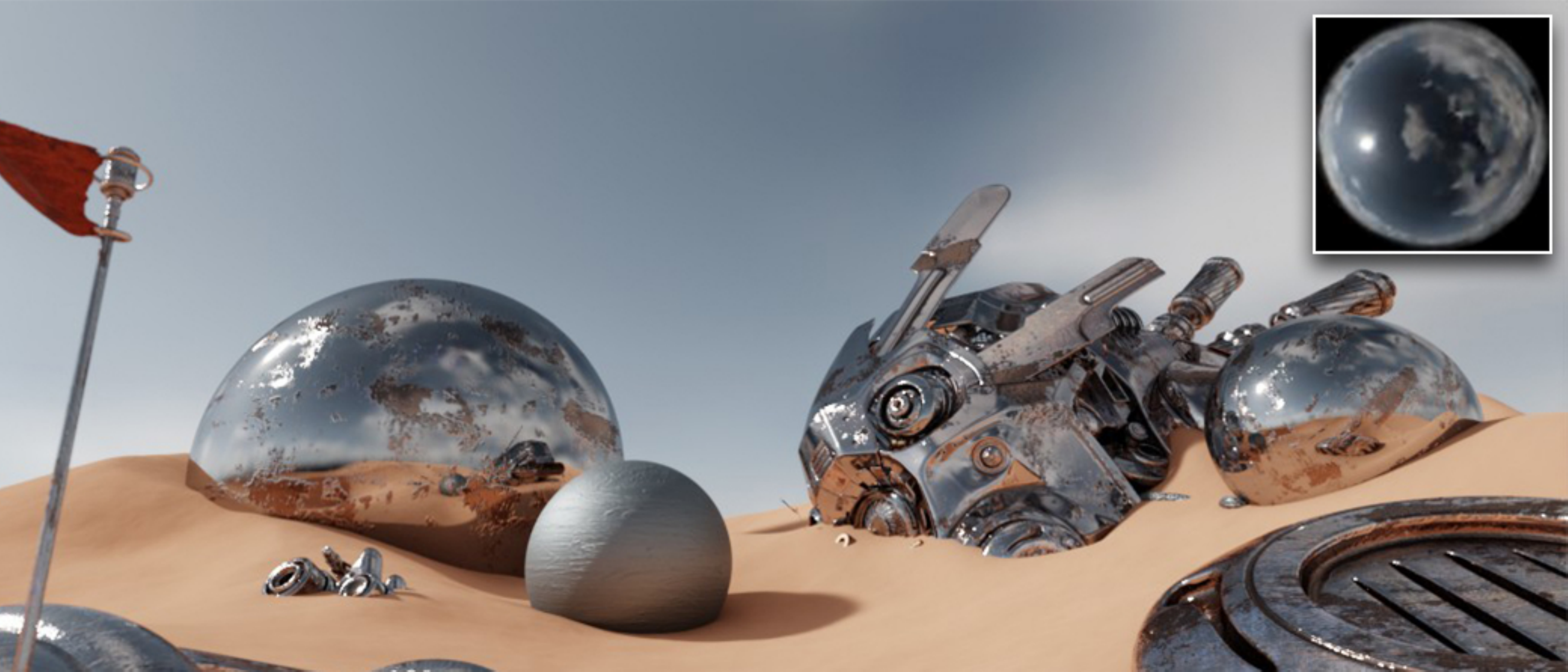}}
          \put(88, 0){\small{\textsf{Input parametric sky}}}
           \put(324, 0){\small{\textsf{Generated photorealistic sky (ours)}}}
          }
    \caption{
    For an input parametric sky (left, inset), our method generates a photorealistic, all-weather HDR environment map (right, inset) while coherently preserving the parametric sky's illumination. This mapping is achieved by combining a generative adversarial pipeline with a differentiable rendering loss. Scene from Blender Studio under CC0 license with added reflective surfaces.
    }
    \label{fig:teaser}
\end{teaserfigure}

\maketitle

\section{Introduction}

Natural illumination is a key component of photorealism, with a pronounced impact on human perception. It is a significant aspect of the design of physical spaces, the planning of urban architecture, and even the perceived visual quality of media \& film. 
The rendering of photorealistic illumination for virtual scenes is generally achieved through the adoption of high dynamic range (HDR) environment maps to guide Image-Based Lighting (IBL) techniques~\cite{ibl}. 
Either physically captured or synthetically generated, HDR imagery (HDRI) is generally understood to represent a hemispherical map of a desired sky dome, projected to a skybox, equirectangular, or other textural representation.
Though conventional low dynamic range (LDR) imagery can be suitable for environment mapping, HDR imaging is integral to the photorealistic rendering of outdoor scenes. 
For physical HDR captures, an estimated $22$ f-stops of exposure are necessary to fully faithfully capture outdoor illumination~\cite{Stumpfel}.
While HDR imaging of real skies is unsurpassed in the photorealism of its output, it is an inflexible solution, as captures are of specific, uncontrolled scene configurations. 

To overcome the rigidity of physical capture, parametric sky models have progressed greatly over the past four decades. 
Originally targeting scientific and engineering applications (e.g., \cite{MOON1940583,perez_model} model only luminance), Nishita et al.~\shortcite{nishita_model} proposed the first color model, with the goal of enabling the rendering of extraterrestrial views of the Earth for space flight simulators.
This model was then extended to include multiple scattering events within the atmosphere, primitive clouds \cite{nishita_model_2}, and later reduced via non-linear least squares fitting by Preetham et al.~\shortcite{preetham_model}.
In order to achieve better accuracy while reducing computational time and memory footprint, these models have evolved to the incumbent Ho{\v{s}}ek-Wilkie-based series of models~\shortcite{hw_model, hw_model_2, hw_model_with_a_vengeance, live_free_or_hw_model}. These are mathematical approximations fitted to data from physical simulations, later validated against real captured skies. While these models show great physical accuracy, they can only properly approximate ideal homogeneous skies, devoid of non-uniform atmospheric formations such as clouds. 

In recent years, the advent of deep learning has lead to the proposal of deep sky models, with the capacity to self-discover features and modalities through latent parameters. 
Notably, this concept has been demonstrated for lighting estimation~\cite{yannick_2017, panonet, skynet, relighting_hdsky}, where LDR images can be used to guide the generation of HDR environment maps for relighting virtual objects and scenes.
Though these learned representations tend to offer little photorealism and generate overly smooth skies, they capture lighting energy with significant accuracy. 
Recently, generative adversarial networks (GANs) targeted towards reproducing clouds have been proposed~\cite{deep_cloud_synthesis, skygan, gan_clouds_thesis}, but these either fail to capture photorealistic textures or lack full HDR capabilities.

In this work, we propose a learned sky model capable of producing photorealistic weathered skies with faithful and controllable HDR illumination. 
Our method combines the versatility of parametric models and the realism of deep generative networks. 
We leverage the Lalonde-Matthews (LM) parametric illumination model~\cite{lm_model}, which represents outdoor skies with parameters such as sun position, sky color, and atmospheric turbidity. This model is also particularly suitable to model overcast skies, having been previously used to guide a state-of-the-art all-weather deep lighting estimation approach~\cite{panonet}.
Given a set of such parameters, we first render a sky dome using the LM equation (see \cref{sec:lm-model-fitting}) and feed the resulting HDR image as input to a deep convolutional network, trained in an adversarial setting on over $25,000$ HDR skydomes from the publicly-available Laval HDR Sky Dataset~\cite{skydb,lm_model}. 
The resulting network, dubbed \textit{LM-GAN}, can be controlled using the same parameters to produce photorealistic weathered HDR skies, which can readily be used in any existing physics-based rendering engine (see \cref{fig:teaser}). 
We demonstrate through quantitative evaluations and ablation studies that our proposed LM-GAN retains parametric fidelity to the LM model and surpasses the current state-of-the-art learning-based models in versatility, fidelity, and photorealism. 
\section{Related work} 

Early models of solar and atmospheric illumination such as \cite{MOON1940583} and \cite{perez_model} were oriented towards scientific and engineering applications. 
With the advent of the digital age, Image-Based Lighting (IBL) techniques~\cite{ibl} proposed the use of High Dynamic Range Imagery (HDRI ~\cite{hdr_book}) to render synthetic objects into real and virtual scenes.
This new paradigm of applications spurred renewed interest in modeling skies and alleviating the burden of physical capture. 

\subsection{Parametric models}

To limit the computational resources required to model sunlight and skylight, it is often preferable to approximate its appearance via a mathematical model and evaluate it against physically captured skies \cite{Kider}.
One such formulation is through the development of numerical models such as ~\cite{nishita_model,nishita_model_2,oneal_model,haber_model,bruneton_model,elek_model}, which derive simplified mathematical representations for complex atmospheric systems. 
Though reducing computation expense, numerical models generally remain complex, memory intensive, and require a pre-computation step.

An alternative formulation introduced by Perez et al.~\shortcite{perez_model} is the fitting of an analytical model to a body of sky data. 
These simpler models can be fitted to computationally expensive, accurate, and diverse data acquired from complex models (e.g., Preetham et al.~\shortcite{preetham_model} is fitted to \cite{nishita_model_2}), path tracers~\cite{hw_model,hw_model_2,hw_model_with_a_vengeance}, or physical simulations such as libRadtran~\cite{libRadtran} (as used in evaluation by Bruneton~\cite{clear_sky_evaluation}).
Such models trade-off accuracy to produce lightweight, flexible, and fast parametric models which support a wide range of applications~\cite{clear_sky_evaluation}.
 
\subsection{Synthetic cloud generation}
Though allowing for versatility through a finite set of intuitive parameters, parametric models are generally limited to clear, hazy, and overcast skies. 
To include more complex atmospheric formations such as clouds, parametric models are often supplemented by volumetric cloud rendering~\cite{DeepScatering} and/or cloud simulations~\cite{clouds_bruneton} through various proposed implementations. 
While this combined approach can be versatile and photorealistic, it can be labor intensive, computationally expensive and difficult to configure \cite{SIGGRAPH_course_2020}.

\subsection{Learned parametric models}

Recently, deep learning methods have been proposed for lighting estimation and the fitting of skies to lighting models. 
Works such as \cite{yannick_2017,panonet} enable HDR re-lighting of virtual objects/scenes by regressing the parameters of parametric lighting models (\cite{hw_model,lm_model}, respectively) from singular, limited field of view LDR images. 
Other approaches~\cite{skynet,Yu2021DualAA,relighting_hdsky} focus on learning the illumination model itself. These methods offer high fidelity for relighting photographs, but do not obtain photorealistic textures, aiming instead at better generalization on the previously-mentioned challenges.

Most recently, approaches have been proposed to augment parametric skies with photorealistic clouds.
Satilmis et al.~\shortcite{deep_cloud_synthesis} proposed a method for augmenting Ho{\v{s}}ek-Wilkie~\shortcite{hw_model} parametric skies per user-controlled cloud placement. This differs from our model in that the user is required to provide a mask positioning the cloud textures, whereas we wish to have clouds naturally positioned as a direct consequence of a learned mapping from parametric inputs to the real world.
SkyGAN~\cite{skygan} proposed fitting the Prague sky model~\shortcite{hw_model_with_a_vengeance} model to real-world photographs, enabling the re-generation of Prague skies with atmospheric formations, though with limited dynamic range, thus requiring an extra virtual sun lamp to be used with rendered scenes. To the best of our knowledge, LM-GAN is the first to produce parametric, photorealistic all-weather HDR skydomes for Image-Based Lighting. 

\begin{figure*}[ht!]
    \includegraphics[width=\textwidth]{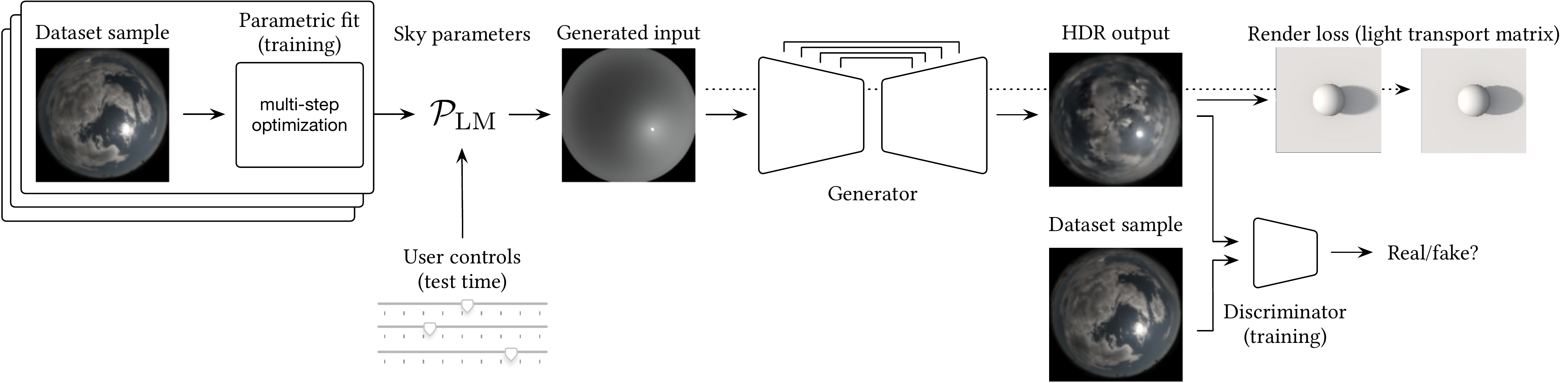}
    \caption{\textbf{Our pipeline.} A generator is trained to convert a parametric input (obtained by fitting a parametric model to a captured HDR sky) to a photorealistic weathered HDR sky via a combination of a patch-based discriminator (GAN) loss, LPIPS (not shown), and a differentiable rendering loss.}
    \label{fig:pipeline}
\end{figure*}

\section{Method}

We briefly summarize our model below and discuss the fitting of the Lalonde-Matthews (LM) model parameters to HDR imagery. In this work, we leverage physically captured images from the publicly-available \textit{Laval HDR Sky Dataset}~\cite{skydb}.

\subsection{Representing real skies parametrically}
\label{sec:lm-model-fitting}

Where previous works such as ~\cite{yannick_2017, skygan} have used the Ho{\v{s}}ek-Wilkie model~\shortcite{hw_model}, we leverage the LM ~\shortcite{lm_model} illumination model instead.  
This model offers 11 parameters for approximating outdoor lighting which can be fit to real HDR skies. It is also less restrictive in its parameters than traditional physically-based models, having thus been shown to be well-suited for learning-based approaches \cite{lm_model,panonet}. 
 
\subsubsection*{The Lalonde-Matthews (LM) model}

The LM model is a hemispherical function of an arbitrary light direction $\mathbf{l}$ (in spherical coordinates) and can be written compactly as the sum of two terms:
\begin{equation}
f_\mathrm{LM}(\mathbf{l}; \: \mathcal{P}_\mathrm{LM}) = 
f_\mathrm{sun}(\mathbf{l}; \: \mathcal{P}_{\mathrm{sun}}, \: \mathbf{l}_{\mathrm{sun}}) +
f_\mathrm{sky}(\mathbf{l}; \: \mathcal{P}_{\mathrm{sky}}, \: \mathbf{l}_{\mathrm{sun}}) \,
\label{eq:lm-model}
\end{equation}
Where $\mathbf{l}_\mathrm{sun} = [\theta_\mathrm{sun}, \varphi_\mathrm{sun}]$ is the sun angular position in spherical coordinates, and $\mathcal{P}_*$ represent term-specific parameters. 
The sky term $f_\mathrm{sky}$ in \cref{eq:lm-model} is the Preetham~\shortcite{preetham_model} sky model $f_\mathrm{Pree}$, parameterized by turbidity $t$ and multiplied channel-wise with an average sky color $\mathbf{c}_\mathrm{sky} \in \mathbb{R}^3$: 
\begin{equation}
f_\mathrm{sky}(\mathbf{l}; \: \mathcal{P}_\mathrm{sky}, \: \mathbf{l}_\mathrm{sun}) = \mathbf{c}_\mathrm{sky} \; f_\mathrm{Pree}(\theta_{\mathrm{sun}}, \: \gamma_{\mathrm{sun}}, \: t) \,
\end{equation}
Where $\gamma_\mathrm{sun}$ is the angle between sky direction $\mathbf{l}$ and the sun $\mathbf{l}_\mathrm{sun}$. 
The sun term $f_\mathrm{sun}$ in \cref{eq:lm-model} is defined as:
\begin{equation}
f_\mathrm{sun}(\mathbf{l}; \: \mathcal{P}_\mathrm{sun}, \: \mathbf{l}_\mathrm{sun}) = \mathbf{c}_\mathrm{sun} \; \exp(-\beta \: \exp(\nicefrac{-\kappa}{\gamma_\mathrm{sun}})) \,
\end{equation}
Where $(\beta, \kappa)$ control the sun scattering, and $\mathbf{c}_\mathrm{sun} \in \mathbb{R}^3$ is the mean sun color. To summarize, the LM model represents the HDR sky dome with the following 11 parameters: 
\begin{equation}
\mathcal{P}_\mathrm{LM} = \left\{\mathbf{c}_\mathrm{sky}, \: t, \: \mathbf{c}_\mathrm{sun}, \: \beta, \: \kappa, \: \mathbf{l}_\mathrm{sun} \right\} \,
\end{equation}

\subsubsection*{LM fitting algorithm}
Similar to \cite{lm_model}, we employ an optimization scheme to fit LM parameters to captured HDR hemispherical skies. 
We assume the sun position $\mathbf{l}_\mathrm{sun}$ can be obtained from capture metadata (geolocation and timestamp) and fine-tuned with a simple local maximum energy search (on patches, to handle overcast scenarios). We then initialize both $\mathbf{c}_\mathrm{sun}$ and $\mathbf{c}_\mathrm{sky}$ to the target sky image's mean color. This is achieved by masking a $30^\circ$ circular region around the sun, sufficient for both sunny and overcast skies. We also exclude suns over $80^\circ$ of zenith, as those skies tend to lack sufficient daylight due to geographic reasons.
Fitting is achieved through the following 4-step strategy: 
\setlist{leftmargin = *} 
\begin{enumerate}[noitemsep,topsep=0pt]

\item A coarse grid-search to initialize scattering parameter $\kappa$. 
In this step, we employ a coarse grid, consisting of $s_{\kappa}$ steps within a range of $[\kappa_{min}, \kappa_{max}]$, combined with standard steps $S$ for both $\beta \in [\beta_{min}, \beta_{max}]$ and $t \in [t_{min}, t_{max}]$. 
We employ two losses: 
\begin{enumerate*}
\item An L1 smoothing loss; between the generated LM sky and target HDR image
\item An L1 rendering loss; between the generated LM sky and target HDR image after transformation by a pre-computed light transport matrix~\cite{light_transport} onto a scene composed of a Lambertian diffuse sphere and ground plane (see \cref{fig:pipeline}).
\end{enumerate*}

\item A fine grid-search to determine scattering parameters $\kappa$, $\beta$, and $t$. The same L1 losses are used. In this step, $\kappa$ is initialized to the best result of the previous step, now with $s_{\kappa}$ steps scaled by $\lambda_\kappa$, in a proportionally smaller range extending in both directions. Parameters $\beta$ and $t$ are iterated as before. The best combination for all $3$ parameters is stored.

\item Optimization of the $\omega_{sun}$ and $\omega_{sky}$ parameters via the Adam optimizer~\cite{adam} with frozen scattering parameters $\kappa$, $\beta$, and $t$ for $n$ iterations. The same L1 losses are used, with added fixed regularization penalties when either the sun or sky color collapses to zero (or an extreme value), and when the color of the sun or sky is neither approximately grayscale, red, or blue (the latter, for skies only).

\item Iterative fine-tuning of all parameters (except sun positioning) with the Adam optimizer, for $n$ more epochs. Regularization and losses are carried forward from Step 3. For reproducibility and details on constants and ranges, see \cref{sec:implementation}.
\end{enumerate}

Using this 4-step strategy, we have annotated over $25,000$ skies from the \textit{Laval HDR Sky Dataset}. 

\subsection{Generating photorealistic skies}

As illustrated in \cref{fig:pipeline}, our generative pipeline takes as input a sky dome (c.f. \cref{sec:lm-model-fitting}), generated from either user-specified parameters or a real sky. 
Then, our model generates a photorealistic weathered sky which preserves the input's HDR illumination. 

\subsubsection*{Preserving HDR illumination}
To preserve the dynamic range, the model's inputs are pre-processed to compressed towards the $[-1,1]$ interval preferred by the model architecture.
This is achieved by
\begin{enumerate*}
    \item re-exposing input skies by dividing by their respective $99^\text{th}$ percentile, 
    \item tonemaping by $I' = \log_2(I + 1)$, and,
    \item shifting inputs towards the $[-1,1]$ interval with $I' = 2I - 1$
\end{enumerate*}. 
Post-generation, this procedure is inversed to recover the dynamic range. 
Through Figure \cref{fig:renders-grid}, we demonstrate this process preserves the dynamic range of the input's illumination. 

\subsubsection*{Network architecture}
The model consists of a UNet~\cite{unet} with fixed update initialization~\cite{fixup} and ReLu activation. 
Dropout~\cite{srivastava2014dropout} is enabled in the bottleneck during training and testing for increased generalization ~\cite{pix2pix}. 
The model halves the resolution and doubles the features at every \textit{down layer}. 
For an input  $s \times s$ image, it has $(log_2(s) - 2)$ \textit{down layers}, ensuring a $4\times4$ resolution at the bottleneck with $2s$ filters. 
Training data is augmented via $x$ and $y$-axes flips and azimuthal rotations, which encourages the network to learn multiple variations of the same sky. 

\subsubsection*{Loss functions}
During training, the network combines three loss functions: patch-based adversarial~\cite{pix2pix}, LPIPS perceptual \cite{lpips}, and a rendering loss. For more details on how each loss contributes to the generated result, see \cref{sec:ablations}. 

The main texture loss is a simple PatchGAN~\cite{pix2pix} loss, directly calculated as an MSE from the output patch feature vectors of a patch-based discriminator with a receptive field of $8\times8$ pixels. The discriminator employs ReLU activations---except for a sigmoid at the end---and instance normalization throughout. Because the goal is to learn general texture and structure, every input given to the discriminator is tonemapped as above.

To further encourage the modeling of details and high frequencies, we also employ an LPIPS~\cite{lpips} perceptual loss of a VGG~\cite{simonyan2014very} network trained on ImageNet~\cite{deng2009imagenet}. This loss has been observed to learn cloud texture details (see \cref{fig:ablation}). As with the generator, this network receives tonemapped images. This loss is scaled so its magnitude is generally $10$ times smaller than the PatchGAN loss, so that it is only given attention during training after the network learned the general structure of the model.

To ensure lighting is preserved, the output of the generator is converted back to linear RGB and used to render a diffuse white sphere on a plane using a light transport matrix (as in \cref{sec:lm-model-fitting}), which is used to calculate a differentiable rendering L1 loss against the same scene lit by the linear LM parametric input. The magnitude of this loss is scaled to be $10$ times smaller than the LPIPS loss, so that emphasis to coherent lighting is only enforced when the visuals are more well-behaved, otherwise HDR artifacts can appear.

\subsection{Implementation details}
\label{sec:implementation}

Our implementation uses PyTorch and was based on UNet FixUp \cite{fixup_unet} and adapted to our generative scenario. The light transport rendering engine and LM rendering engine, as well as the LM fitting algorithm, are all PyTorch scripts vectorized and with GPGPU. Models were trained until convergence (around $440$ epochs), with batch sizes chosen as the highest value to fit the GPU used for training.
On a single Titan X GPU, the $128 \times 128$ models took on average 15 minutes per epoch to train and consuming 11 GB of VRAM on a batch size of $32$. 
For the fitting algorithm, we employed the search ranges for $(\kappa, \beta, t)$ as $[0, 1], [0, 50], [2, 20]$, respectively, determined empirically by finding points where the model became incoherent. Our search used $s_{\kappa}=\lambda_\kappa=0.1$, with $S=2$ and $n=1000$.
\section{Experiments}
\label{sec:results}

We compare our technique to the state of the art, both in terms of sky appearance and HDR lighting fidelity. We also evaluate how well our parametric fitting compares to state-of-the-art fitting approaches.

\begin{figure*}
    \centering
    \setlength{\tabcolsep}{1pt}
    \renewcommand{\arraystretch}{0.5}
    \newlength{\tmplength}
    \setlength{\tmplength}{0.135\linewidth}
    \begin{tabular}{cccccccc}
    & \small{\textsf{HW~\shortcite{hw_model}}} & \small{\textsf{Prague~\shortcite{hw_model_with_a_vengeance}}} & \small{\textsf{LM~\shortcite{lm_model}}} & \small{\textsf{SkyNet~\shortcite{skynet}}} & \small{\textsf{SkyGAN~\shortcite{skygan}}} & \small{\textsf{LM-GAN (ours)}} & \small{\textsf{reference}} \\ 
    \rotatebox{90}{\parbox{2.3cm}{\centering \small{\textsf{sunrise/sunset}}}} & 
    \includegraphics[width=\tmplength]{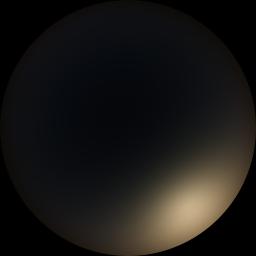} & 
    \includegraphics[width=\tmplength]{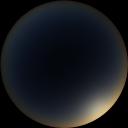} & 
    \includegraphics[width=\tmplength]{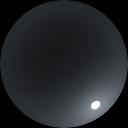} & 
    \includegraphics[width=\tmplength]{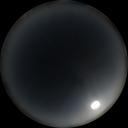} & 
    \includegraphics[width=\tmplength]{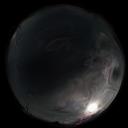} & 
    \includegraphics[width=\tmplength]{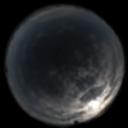} & 
    \includegraphics[width=\tmplength]{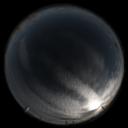} \\ 
    \rotatebox{90}{\parbox{2.3cm}{\centering \small{\textsf{sunny}}}} & 
    \includegraphics[width=\tmplength]{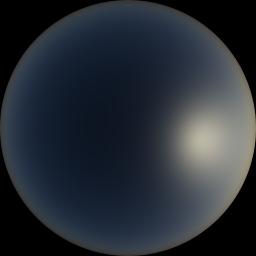} & 
    \includegraphics[width=\tmplength]{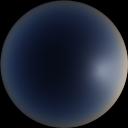} & 
    \includegraphics[width=\tmplength]{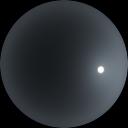} & 
    \includegraphics[width=\tmplength]{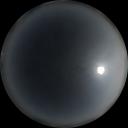} & 
    \includegraphics[width=\tmplength]{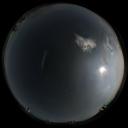} & 
    \includegraphics[width=\tmplength]{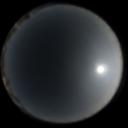} & 
    \includegraphics[width=\tmplength]{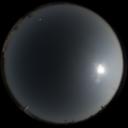} \\ 
    \rotatebox{90}{\parbox{2.3cm}{\centering \small{\textsf{mostly sunny}}}} & 
    \includegraphics[width=\tmplength]{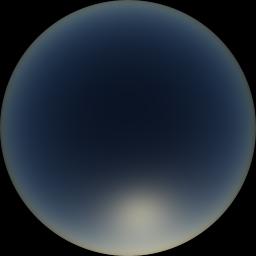} & 
    \includegraphics[width=\tmplength]{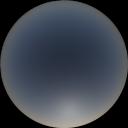} & 
    \includegraphics[width=\tmplength]{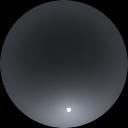} & 
    \includegraphics[width=\tmplength]{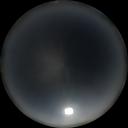} & 
    \includegraphics[width=\tmplength]{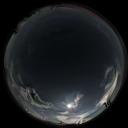} & 
    \includegraphics[width=\tmplength]{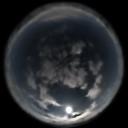} & 
    \includegraphics[width=\tmplength]{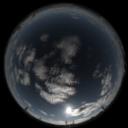} \\ 
    \rotatebox{90}{\parbox{2.3cm}{\centering \small{\textsf{partly cloudy}}}} & 
    \includegraphics[width=\tmplength]{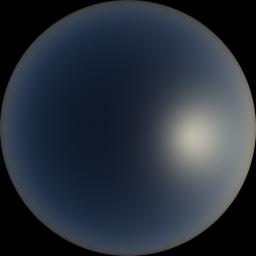} & 
    \includegraphics[width=\tmplength]{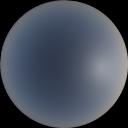} & 
    \includegraphics[width=\tmplength]{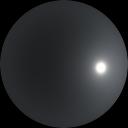} & 
    \includegraphics[width=\tmplength]{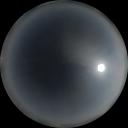} & 
    \includegraphics[width=\tmplength]{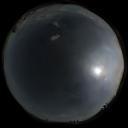} & 
    \includegraphics[width=\tmplength]{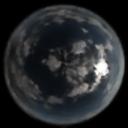} & 
    \includegraphics[width=\tmplength]{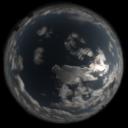} \\ 
    \rotatebox{90}{\parbox{2.3cm}{\centering \small{\textsf{mostly cloudy}}}} & 
    \includegraphics[width=\tmplength]{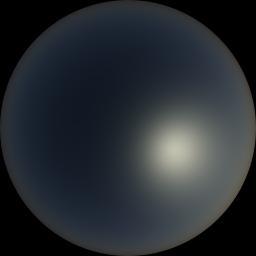} & 
    \includegraphics[width=\tmplength]{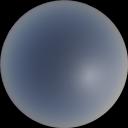} & 
    \includegraphics[width=\tmplength]{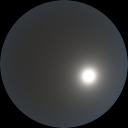} & 
    \includegraphics[width=\tmplength]{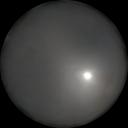} & 
    \includegraphics[width=\tmplength]{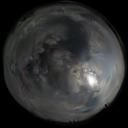} & 
    \includegraphics[width=\tmplength]{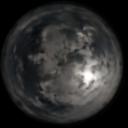} & 
    \includegraphics[width=\tmplength]{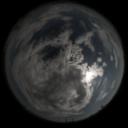} \\ 
    \rotatebox{90}{\parbox{2.3cm}{\centering \small{\textsf{overcast}}}} & 
    \includegraphics[width=\tmplength]{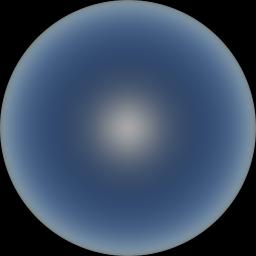} & 
    \includegraphics[width=\tmplength]{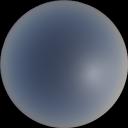} & 
    \includegraphics[width=\tmplength]{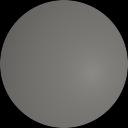} & 
    \includegraphics[width=\tmplength]{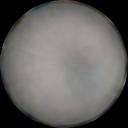} & 
    \includegraphics[width=\tmplength]{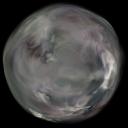} & 
    \includegraphics[width=\tmplength]{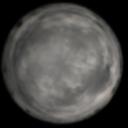} & 
    \includegraphics[width=\tmplength]{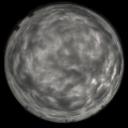} \\ 
    \end{tabular}    
    \caption{\textbf{Visual comparison of different sky lighting models}. While physics-based parametric models (\cite{hw_model}, \cite{hw_model_with_a_vengeance}) are physically accurate, they fail to accurately represent skies with diverse atmostpheric conditions, such as clouds. The LM~\shortcite{lm_model} and SkyNet~\shortcite{skynet} models are more versatile but yield oversmooth results. SkyGAN~\shortcite{skygan} generates somewhat unrealistic clouds. In contrast, our approach maintains the versatility of parametric models and the photorealism of photographic reference. }
    \label{fig:visuals-grid}
\end{figure*}

\subsection{Methodology}
\label{sec:methodology}

\subsubsection*{Dataset} We employ the Laval HDR Skies dataset~\cite{skydb} to evaluate all methods, from which we sample a test set that was not accessible during training (c.f. \cref{sec:lm-model-fitting}). For evaluation we bin this test set by NOAA forecast weather descriptions~\cite{noaa}, first using color-based binary thresholding~\cite{heinle2010automatic}. By counting the percentage of cloud coverage, we classify the test set as follows: $0$--$\nicefrac{1}{8}$ of the sky covered by clouds is named a \textit{sunny} sky (su), $\nicefrac{1}{8}$--$\nicefrac{3}{8}$ \textit{mostly sunny} (ms), $\nicefrac{3}{8}$--$\nicefrac{5}{8}$ \textit{partly cloudy} (pc), $\nicefrac{5}{8}$--$\nicefrac{7}{8}$ \textit{mostly cloudy} (mc), and $\nicefrac{7}{8}$--$1$ \textit{overcast} (oc). Moreover, we also separate the skies where the sun is within $20^\circ$ of the horizon as a \textit{sunrise/sunset} (ss) category, due to their different color and lighting configurations. To ensure the train and test sets do not share time-adjacent frames, the test set is composed of completely separate days of capture, never seen in training. Each day in the test set is from approximately a different month, resulting in $2,580$ samples completely separate from the dataset's remaining $23,000$. The final distribution per bin in the test set is $565$ sunny, $534$ mostly sunny, $380$ partly cloudy, $375$ mostly cloudy, $396$ overcast, $330$ sunrise/sunset.

\subsubsection*{Metrics} To evaluate the appearance and diversity of sky models, we use the FID measure~\cite{heusel2017gans} to compare their distributions of appearance. Prior to computing the FID, skies are re-exposed to their $99^{th}$ percentile, then applied a $\gamma=2.2$ tone mapping and finally clamped to a range between $0$ and $1$.
We further consider the RMSE metric and its scale-invariant counterpart, si-RMSE~\cite{evaluations4images}, computed on the generated skies. The former captures the absolute amount of energy present in the sky, penalizing a lack of strong sun values, while the latter compensates for intensity mismatches and focuses on correct contrast. We compute the RMSE/si-RMSE scores on a Lambertian diffuse scene rendered with a precomputed transport matrix (c.f. \cref{sec:lm-model-fitting}).

\subsubsection*{Parametric models} We first consider three parametric sky models: the combined sun and sky Ho\v{s}ek-Wilkie~\shortcite{hw_model_2} (fit using the algorithm of \cite{yannick_2017}); its extension, the ``Prague Sky Model''~\cite{hw_model_with_a_vengeance} (fit using the algorithm of \cite{skygan}; and the Lalonde-Matthews~\shortcite{lm_model} (fit by our custom algorithm, see~\cref{sec:lm-model-fitting}). All parametric models are optimized to our HDR test set. 

\subsubsection*{Learned models} We also compare against the deep sky model SkyNet \cite{skynet} trained on the same dataset as our model; and the current state-of-the-art neural sky model, SkyGAN~\cite{skygan}, which we re-implemented from the information provided in the paper and also trained on the same dataset as ours, using the aforementioned Ho\v{s}ek-Wilkie fits as input.

SkyGAN relies on StyleGAN3~\cite{stylegan3} and also produces a realistic sky given a parametric input. However, SkyGAN produces a residual, which is \emph{added} to the parametric input. We also observe, as in \cite{skygan}, that it is not able to preserve HDR lighting from the input. 

\begin{table}[t]
    \setlength\tabcolsep{2pt}
    \centering
    \small
    \begin{tabular}{clccccccc}
        \toprule
        & model & \textbf{\textsf{su}} & \textbf{\textsf{ms}} & \textbf{\textsf{pc}} & \textbf{\textsf{mc}} & \textbf{\textsf{oc}} & \textbf{\textsf{ss}} & \textbf{\textsf{all}}\\
        \midrule
        \multirow{6}{*}{\rotatebox{90}{\textsf{FID}}} & 
         \textsf{\textbf{HW}~\shortcite{hw_model_2}}                    & $142.83$ & $200.17$ & $208.81$ & $202.53$ & $175.36$ & $188.08$ & $175.29$ \\
        & \textsf{\textbf{Prague}~\shortcite{hw_model_with_a_vengeance}} & $170.32$ & $228.89$ & $233.84$ & $224.49$ & $195.52$ & $207.79$ & $202.46$ \\
        &  \textsf{\textbf{LM}~\shortcite{lm_model}}                      & $147.13$ & $190.72$ & $205.27$ & $197.36$ & $174.02$ & $178.06$ & $169.95$ \\
        &  \textsf{\textbf{SkyNet}~\shortcite{skynet}}                    & $126.24$ & $187.22$ & $209.72$ & $214.07$ & $190.65$ & $167.61$ & $165.83$ \\
        & \textsf{\textbf{SkyGAN}~\shortcite{skygan}}                    & $40.95$ & $38.94$ & $46.27$ & $55.12$ & $71.23$ & $73.96$ & $34.89$ \\
        & \textsf{\textbf{LM-GAN} (ours)}                                & $\mathbf{29.57}$ & $\mathbf{28.03}$ & $\mathbf{29.15}$ & $\mathbf{30.38}$ & $\mathbf{38.59}$ & $\mathbf{38.46}$ & $\mathbf{20.28}$ \\
        \midrule
        \multirow{6}{*}{\rotatebox{90}{\textsf{RMSE}}} & 
        \textsf{\textbf{HW}~\shortcite{hw_model_2}}                    & $29.02$ & $19.74$ & $12.78$ & $5.75$ & $2.89$ & $7.89$  &  $14.61$\\
        & \textsf{\textbf{Prague}~\shortcite{hw_model_with_a_vengeance}} & $29.02$ & $19.75$ & $12.81$ & $5.79$ & $2.92$ & $7.9$ &  $14.63$\\
        & \textsf{\textbf{LM}~\shortcite{lm_model}}                      & $22.48$ & $13.25$ & $9.36$ & $5.74$ & $3.83$ & $7.66$ & $11.45$ \\
        & \textsf{\textbf{SkyNet}~\shortcite{skynet}}                    & $26.62$ & $18.24$ & $12.79$ & $6.27$ & $\mathbf{2.71}$ & $7.84$ & $13.82$ \\
        & \textsf{\textbf{SkyGAN}~\shortcite{skygan}}                    & $28.98$ & $19.72$ & $12.77$ & $5.73$ & $2.86$ & $7.86$ & $14.59$ \\
        & \textsf{\textbf{LM-GAN} (ours)}                                & $\mathbf{21.71}$ & $\mathbf{12.56}$ & $\mathbf{8.70}$ & $\mathbf{4.97}$ & $3.16$ & $\mathbf{7.18}$ & $\mathbf{10.76}$ \\
        \midrule
        \multirow{6}{*}{\rotatebox{90}{\textsf{si-RMSE}}} & 
        \textsf{\textbf{HW}~\shortcite{hw_model_2}}                    & $28.99$ & $19.72$ & $12.76$ & $5.73$ & $2.87$ & $7.86$ & $14.59$ \\
        & \textsf{\textbf{Prague}~\shortcite{hw_model_with_a_vengeance}} & $29.02$ & $19.74$ & $12.78$ & $5.75$ & $2.89$ & $7.89$ & $14.62$ \\
        & \textsf{\textbf{LM}~\shortcite{lm_model}}                      & $20.18$ & $11.43$ & $7.50$ & $3.31$ & $1.43$ & $6.59$  & $9.43$  \\
        & \textsf{\textbf{SkyNet}~\shortcite{skynet}}                    & $25.08$ & $16.76$ & $10.96$ & $5.01$ & $2.43$ & $7.02$ & $12.58$ \\
        & \textsf{\textbf{SkyGAN}~\shortcite{skygan}}                    & $28.64$ & $19.57$ & $12.72$ & $5.71$ & $2.84$ & $7.66$ & $14.44$ \\
        & \textsf{\textbf{LM-GAN} (ours)}                                & $\mathbf{19.72}$ & $\mathbf{11.08}$ & $\mathbf{7.30}$ & $\mathbf{3.16}$ & $\mathbf{1.37}$ & $\mathbf{6.31}$ & $\mathbf{9.17}$ \\
        \bottomrule
    \end{tabular}
    \caption{\textbf{Error on the sky texture.} The columns indicate the splits according to cloudiness (c.f. \cref{sec:methodology}), ranging from sunny (su) to overcast (oc), and sunrise/sunset (ss). ``all'' indicates metrics averaged over the entire test set.}
    \label{tab:texture}
    \vspace{-5mm}
\end{table}
\begin{table}[tbp]
    \setlength\tabcolsep{2pt}
    \centering
    \footnotesize
    \renewcommand{\arraystretch}{1.2}
    \begin{tabular}{@{}clccccccc@{}}
        \toprule
        & model & \textbf{\textsf{su}} & \textbf{\textsf{ms}} & \textbf{\textsf{pc}} & \textbf{\textsf{mc}} & \textbf{\textsf{oc}} & \textbf{\textsf{ss}} & \textbf{\textsf{all}}\\
        \midrule
        \multirow{6}{*}{\rotatebox{90}{\scriptsize\textsf{RMSE (ref.)}}} & 
        \textsf{\textbf{HW}~\shortcite{hw_model_2}}                    & $6417.78$ & $4431.04$ & $3138.49$ & $2193.35$ & $1898.95$ & $1529.03$ & $3591.46$ \\
        & \textsf{\textbf{Prague}~\shortcite{hw_model_with_a_vengeance}} & $4767.83$ & $3180.79$ & $2348.25$ & $1484.03$ & $1248.14$ & $1098.06$ & $2596.64$ \\
        & \textsf{\textbf{LM}~\shortcite{lm_model}}                      & $\mathbf{1163.16}$ & $716.64$ & $582.67$ & $407.78$ & $320.21$ & $\mathbf{120.11}$ & $612.84$ \\
        & \textsf{\textbf{SkyNet}~\shortcite{skynet}}                    & $2596.66$ & $2236.56$ & $2276.70$ & $1562.40$ & $970.81$ & $995.53$ & $1870.67$ \\
        & \textsf{\textbf{SkyGAN}~\shortcite{skygan}}                    & $6235.02$ & $4375.28$ & $2959.59$ & $1858.38$ & $1432.77$ & $1448.21$ & $3382.92$ \\
        & \textsf{\textbf{LM-GAN} (ours)}                                & $1175.01$ & $\mathbf{707.38}$ & $\mathbf{571.89}$ & $\mathbf{402.45}$ & $\mathbf{308.04}$ & $125.21$ & $\mathbf{609.94}$ \\
        \midrule
        \multirow{6}{*}{\rotatebox{90}{\scriptsize\textsf{si-RMSE (ref.)}}} & 
        \textsf{\textbf{HW}~\shortcite{hw_model_2}}                     & $2158.68$ & $1591.31$ & $1078.51$ & $742.31$ & $825.80$ & $547.31$ & $1265.88$ \\
        & \textsf{\textbf{Prague}~\shortcite{hw_model_with_a_vengeance}} & $1958.56$ & $1461.44$ & $1150.64$ & $918.86$ & $895.20$ & $509.38$ & $1237.26$ \\
        & \textsf{\textbf{LM}~\shortcite{lm_model}}                      & $\mathbf{158.29}$ & $\mathbf{113.75}$ & $\mathbf{107.58}$ & $117.70$ & $120.09$ & $66.46$ & $\mathbf{118.12}$ \\
        & \textsf{\textbf{SkyNet}~\shortcite{skynet}}                   & $311.97$ & $284.94$ & $238.47$ & $196.44$ & $144.37$ & $177.05$ & $235.80$ \\
        & \textsf{\textbf{SkyGAN}~\shortcite{skygan}}                    & $1278.86$ & $870.70$ & $529.37$ & $285.27$ & $224.39$ & $363.76$ & $660.79$ \\
        & \textsf{\textbf{LM-GAN} (ours)}                                & $171.05$ & $121.78$ & $112.15$ & $\mathbf{117.28}$ & $\mathbf{115.81}$ & $\mathbf{62.82}$ & $122.06$ \\
        \midrule
        \multirow{2}{*}{\rotatebox{90}{\scriptsize\textsf{RMSE}}} & 
        \textsf{\textbf{SkyGAN}~\shortcite{skygan}}                    & $506.93$ & $526.74$ & $606.77$ & $742.99$ & $1028.20$ & $574.81$ & $648.77$ \\
        & \textsf{\textbf{LM-GAN} (ours)}                               & $\mathbf{234.16}$ & $\mathbf{170.08}$ & $\mathbf{139.93}$ & $\mathbf{104.94}$ & $\mathbf{84.46}$ & $\mathbf{55.40}$ & $\mathbf{142.43}$ \\
        \midrule
        \multirow{2}{*}{\rotatebox{90}{\scriptsize\textsf{si-RMSE}}} & 
        \textsf{\textbf{SkyGAN}~\shortcite{skygan}}                    & $319.59$ & $329.63$ & $360.12$ & $381.01$ & $484.37$ & $313.51$ & $361.10$ \\
        & \textsf{\textbf{LM-GAN} (ours)}                              & $\mathbf{95.53}$ & $\mathbf{78.62}$ & $\mathbf{69.46}$ & $\mathbf{58.94}$ & $\mathbf{53.88}$ & $\mathbf{42.03}$ & $\mathbf{69.65}$ \\
        \noalign{\vskip 1mm}
        \bottomrule
    \end{tabular}
    \caption{\textbf{Lighting error on rendered scenes.} The columns indicate the splits according to cloudiness (c.f. \cref{sec:methodology}), ranging from sunny (su) to overcast (oc), and sunrise/sunset (ss). ``all'' indicates the average over the entire test set.}
    \label{tab:lighting}
    \vspace{-5mm}
\end{table}

\subsection{HDR sky synthesis}

\subsubsection*{Quantitative evaluation}

We first compare all techniques on our HDR environment map test set in skyangular format in \cref{tab:texture}. We observe that increasing the cloud coverage significantly worsens the FID of parametric sky models---attaining an FID of over $200$. In comparison, our method achieves consistent robustness, down to a value of $20.28$ on the entire test set's distribution, and $38.46$ in the worst case. Overall, the RMSE measure tells a different story, where increasing cloud coverage provides lower error, as the amount of energy present in the environment map lowers drastically with clouds occluding the sun. 

We can further see the impact of the energy on renders in \cref{tab:lighting}. Similarly to the previous experiment, our method achieves state-of-the-art results. Interestingly, in some categories our method improved upon the results of the LM environment map it received as input, showcasing an advantage in correlating lighting and texture for more scattered illumination configurations. In this table it is also worth highlighting the advantage of LM model's flexibility, allowing a much more precise parametric fit to the ground truth panoramas.
In the first part of the table, all the measures are computed with respect to the reference ground truth environment map from our test set. 
Contrarily to parametric models, neither SkyGAN nor our method intend to represent a captured environment map, but instead to preserve the appearance and energy present in their parametric input. As such, the limited expressivity of their parametric input acts as an upper bound for the accuracy of these methods, which we believe to be a small price to pay against the user editability these inputs provide. The second part of \cref{tab:lighting} presents a comparison of renders against each method's input, to present more faithfully the capabilities of both methods to preserve the environment map energy. SkyGAN presented visible improvements for clearer skies, possibly due to the parametric model used as input combined with its task of reconstructing the input, but this improvement was not consistent in cloudier skies. Our model improved considerably as more clouds covered the sun, showing again a successful mapping from the clear skies parametric input to all-weather skies.

\begin{figure*}
    \centering
    \footnotesize
    \setlength{\tabcolsep}{1pt}
    \renewcommand{\arraystretch}{0.8}
    \setlength{\tmplength}{0.13\linewidth}
    \begin{tabular}{cccccccc}
    & \small{\textsf{HW~\shortcite{hw_model}}} & \textsf{Prague~\shortcite{hw_model_with_a_vengeance}} & {\textsf{LM~\shortcite{lm_model}}} & \small{\textsf{SkyNet~\shortcite{skynet}}} & \textsf{SkyGAN~\shortcite{skygan}} & {\textsf{LM-GAN (ours)}} & {\textsf{reference}} \\ 
    \rotatebox{90}{\parbox{1.5cm}{\centering {\textsf{sunrise/sunset}}}} & 
    \includegraphics[width=\tmplength]{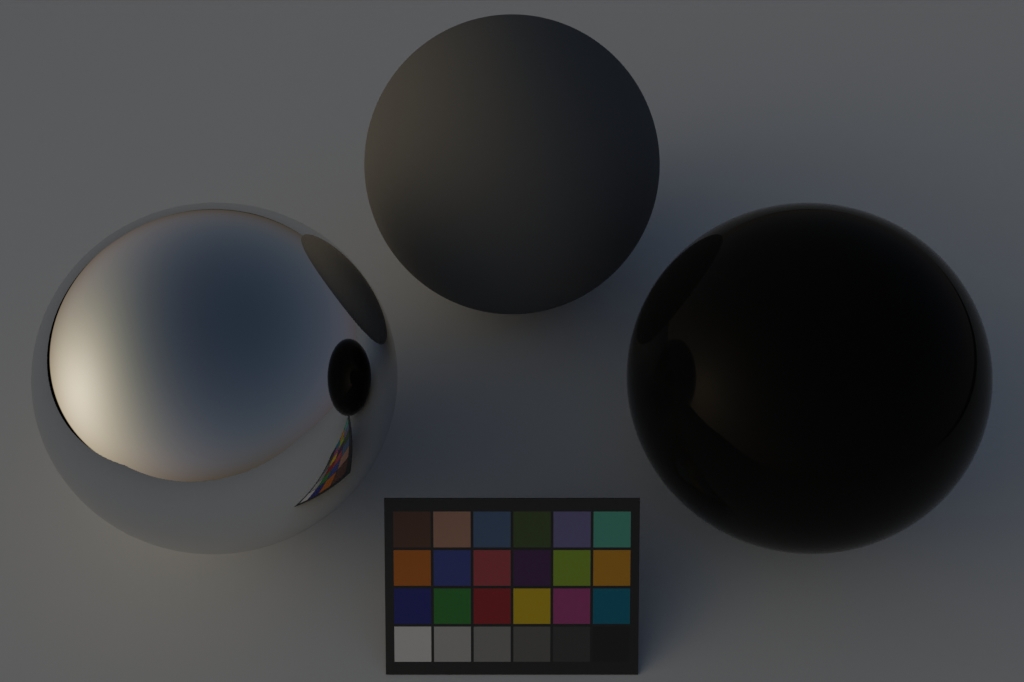} & 
    \includegraphics[width=\tmplength]{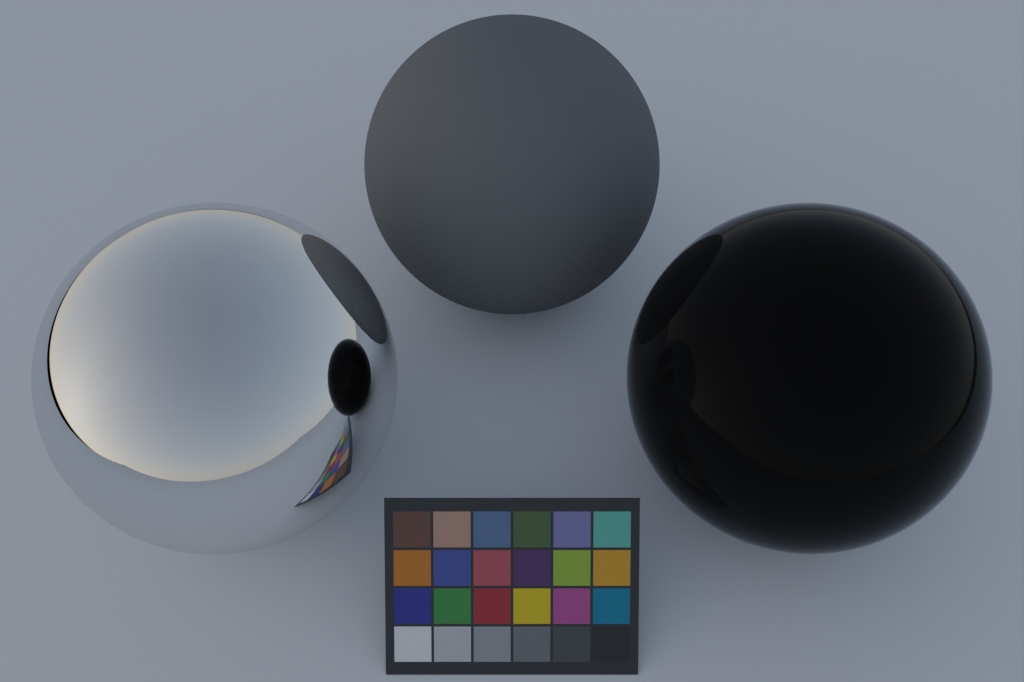} & 
    \includegraphics[width=\tmplength]{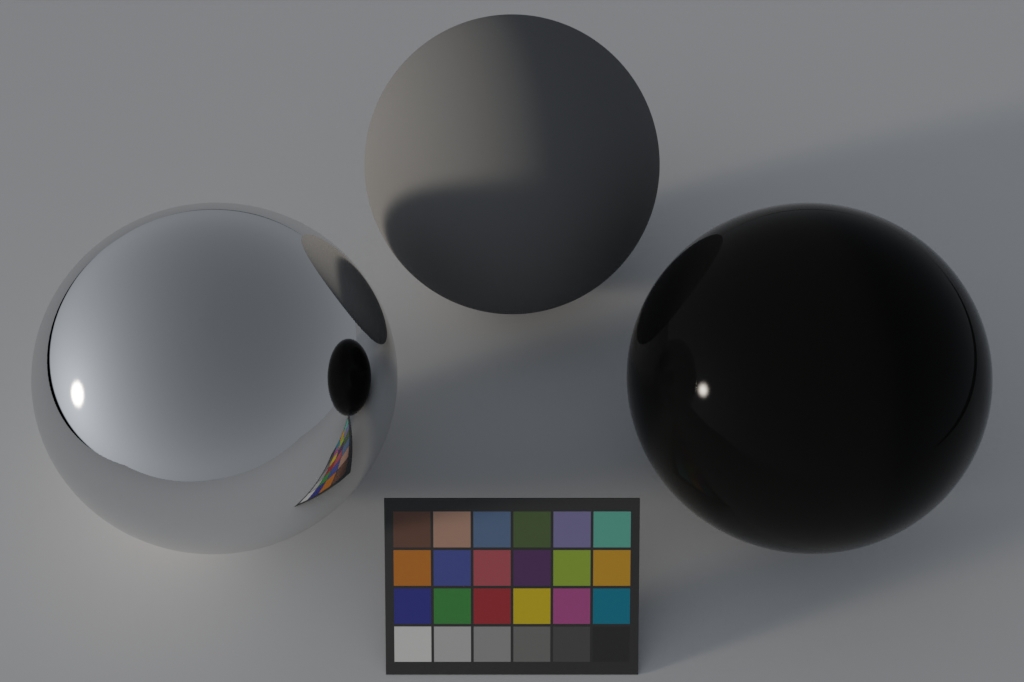} & 
    \includegraphics[width=\tmplength]{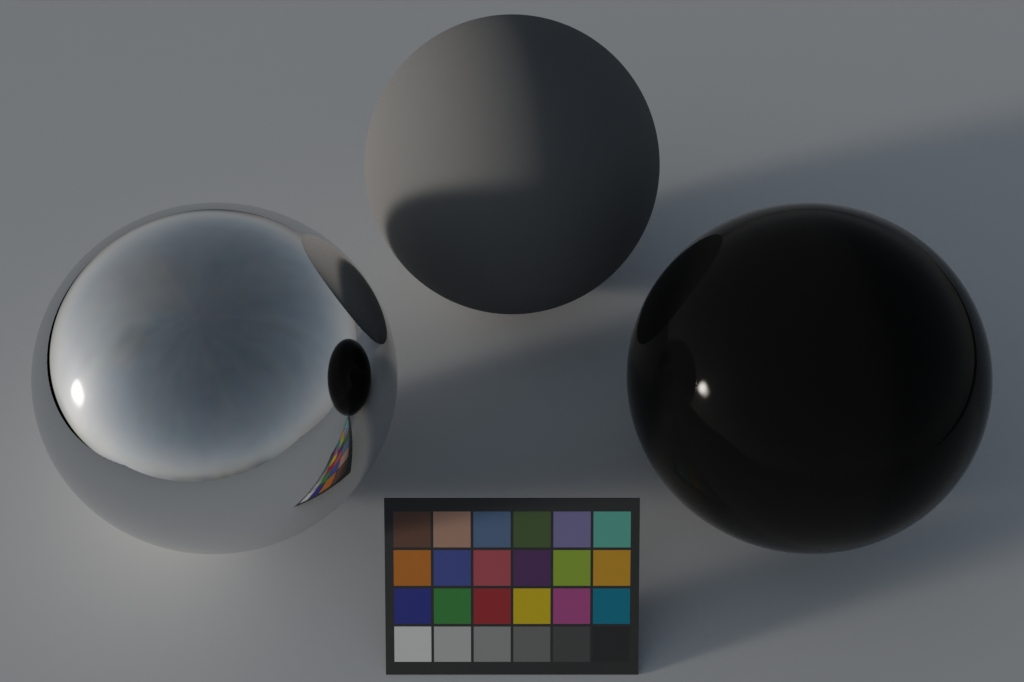} & 
    \includegraphics[width=\tmplength]{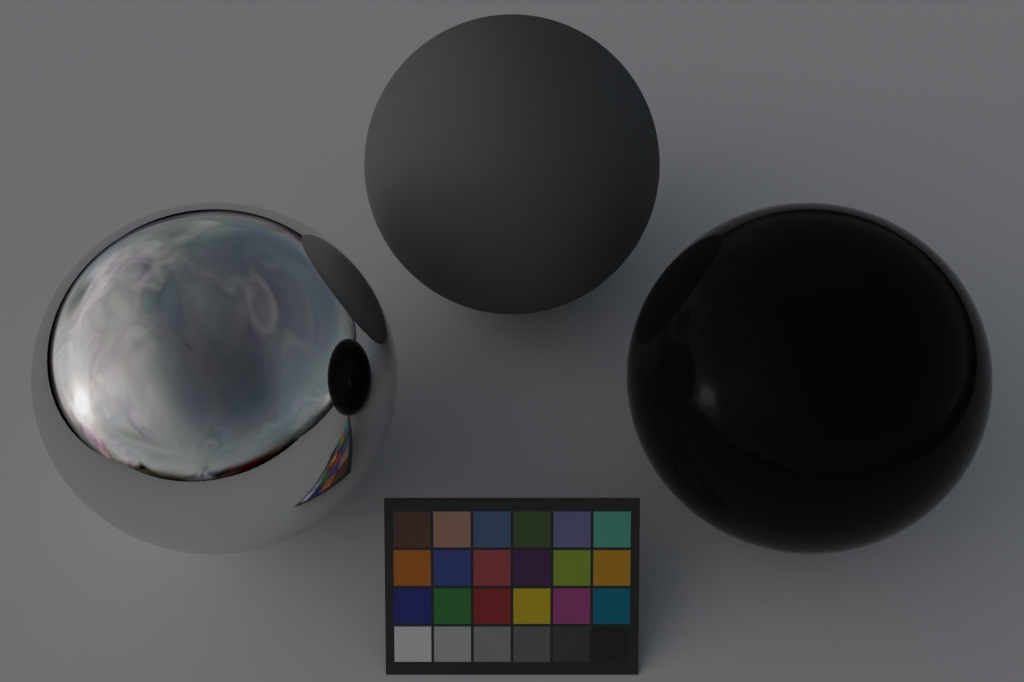} & 
    \includegraphics[width=\tmplength]{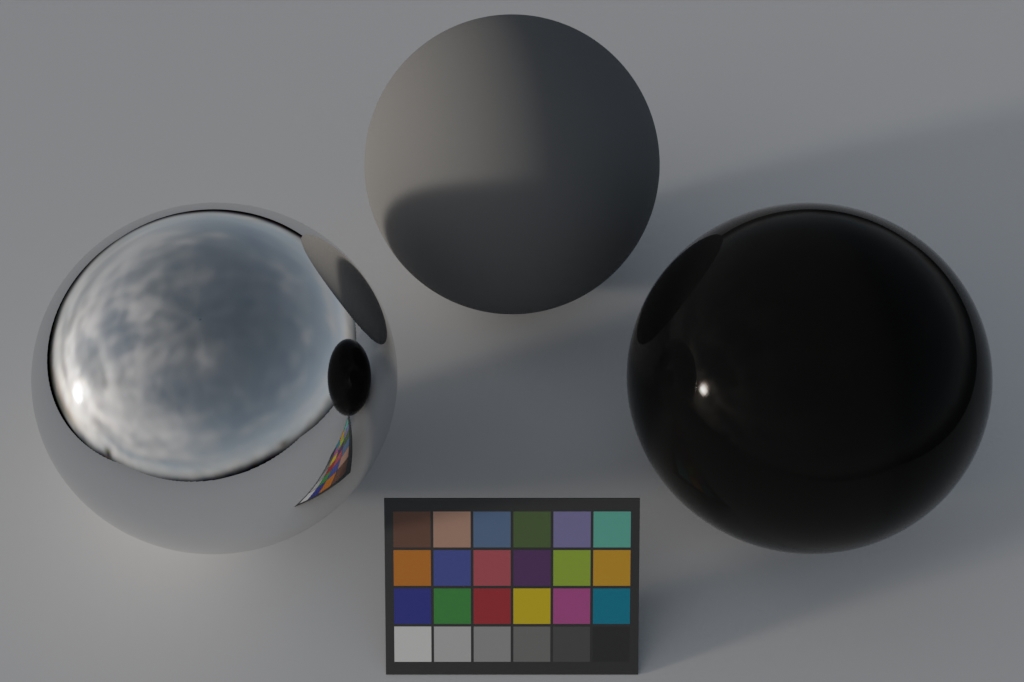} & 
    \includegraphics[width=\tmplength]{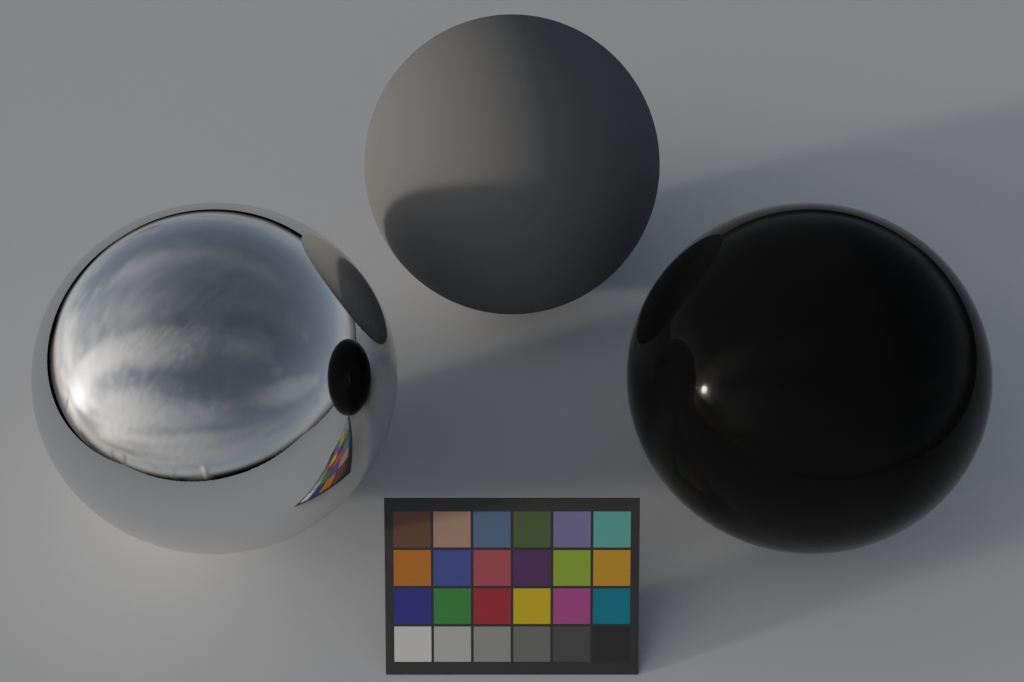} \\ 
    \rotatebox{90}{\parbox{1.5cm}{\centering {\textsf{sunny}}}} & 
    \includegraphics[width=\tmplength]{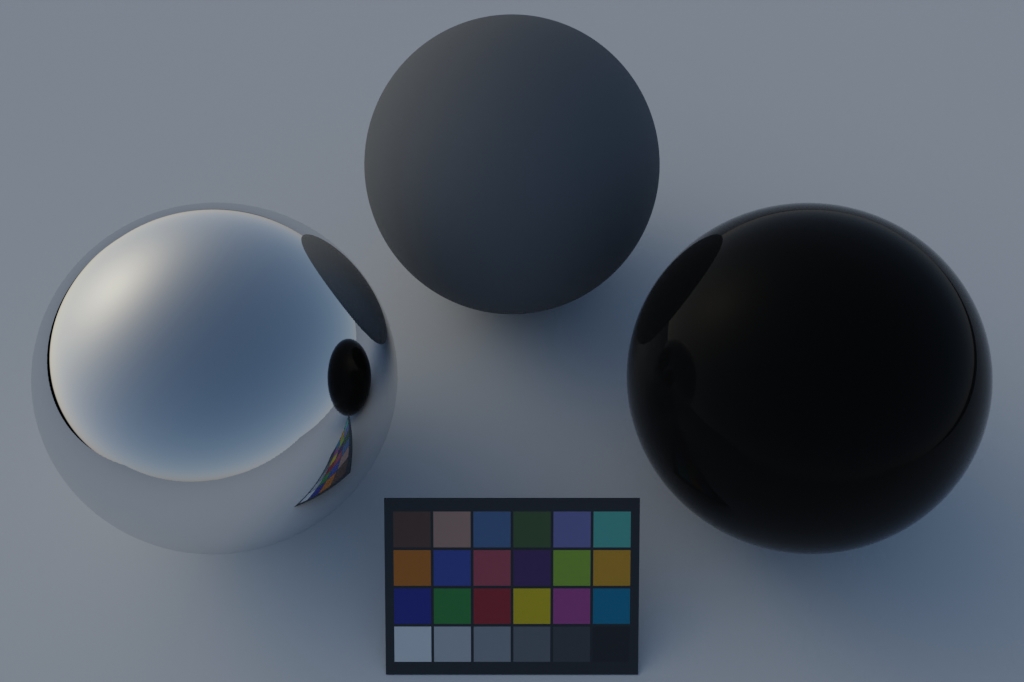} & 
    \includegraphics[width=\tmplength]{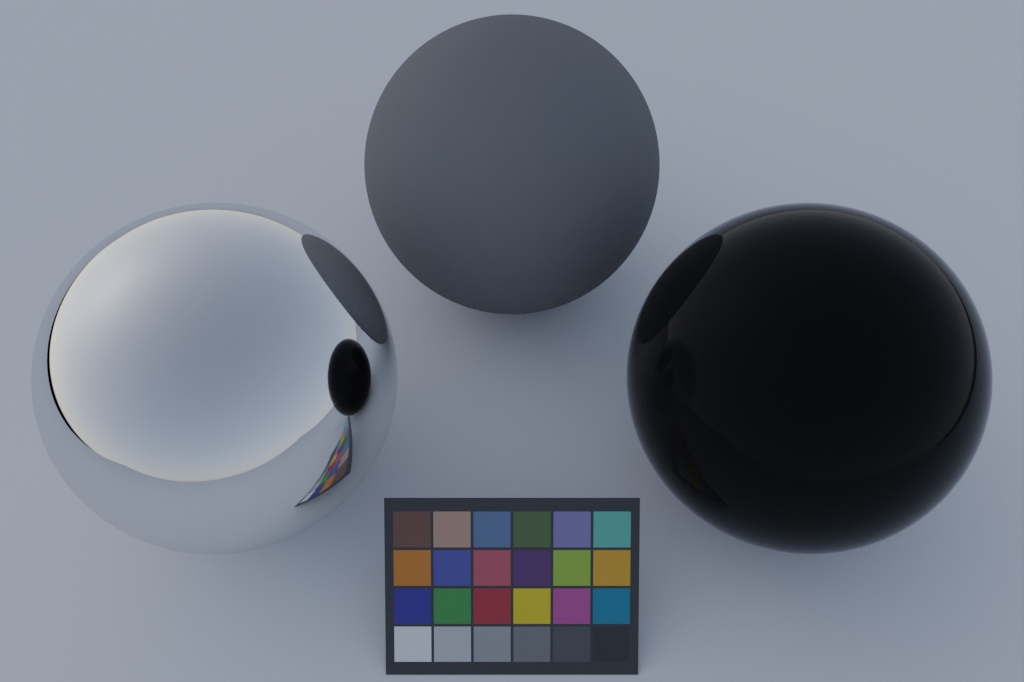} & 
    \includegraphics[width=\tmplength]{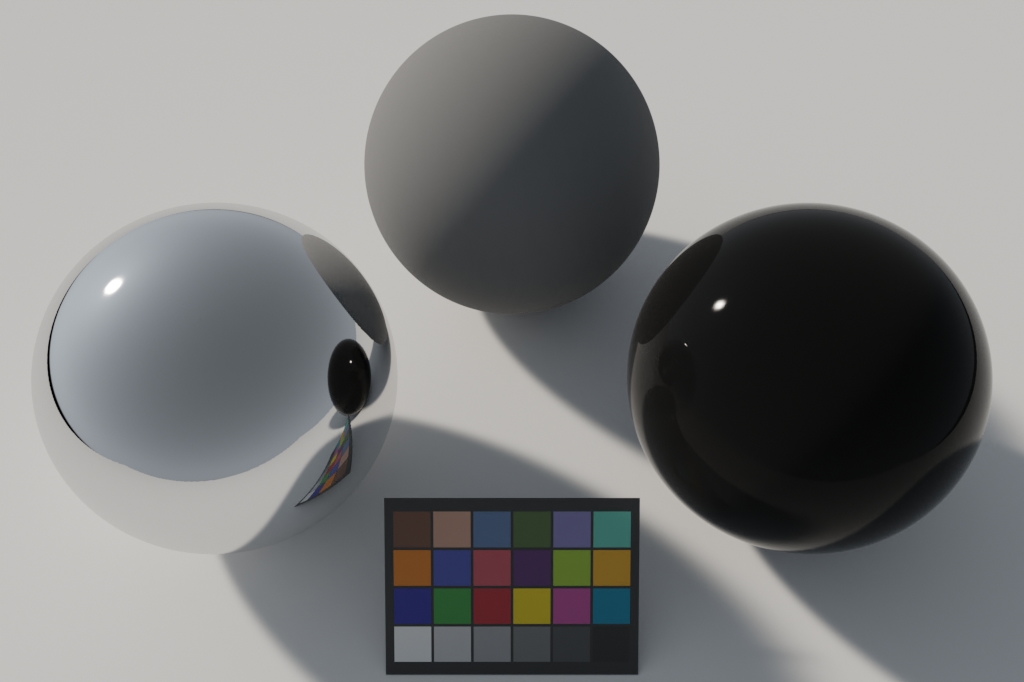} & 
    \includegraphics[width=\tmplength]{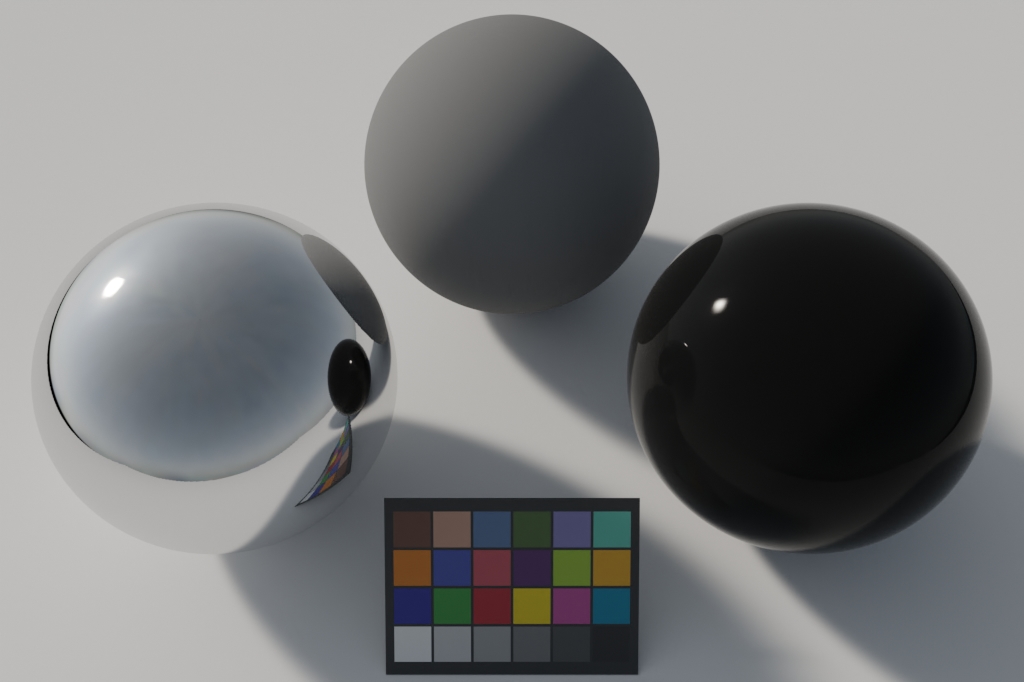} & 
    \includegraphics[width=\tmplength]{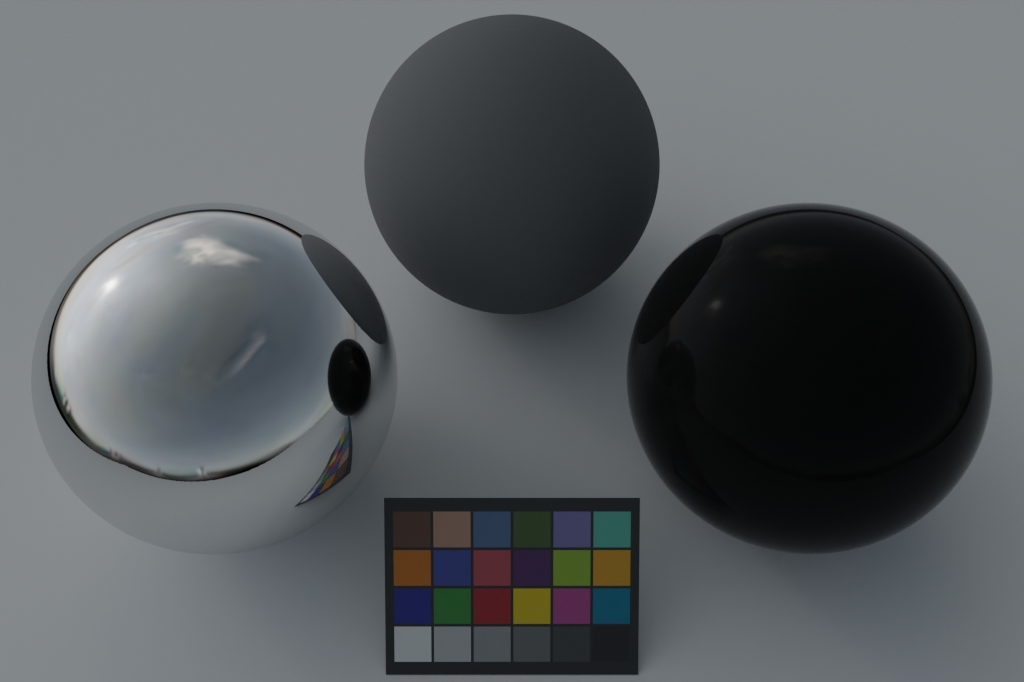} & 
    \includegraphics[width=\tmplength]{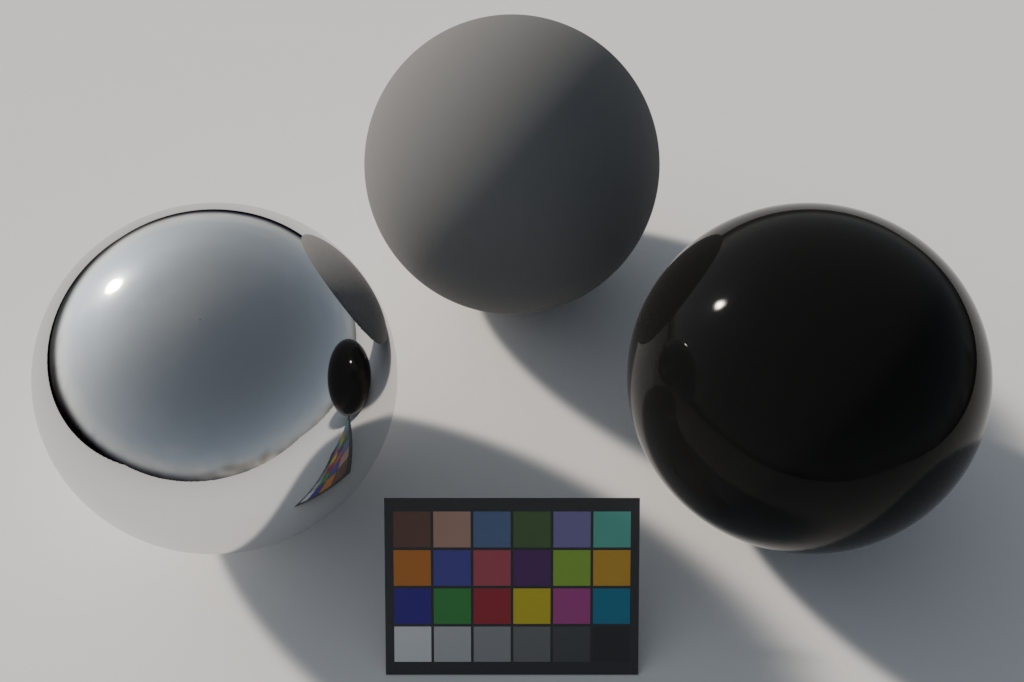} & 
    \includegraphics[width=\tmplength]{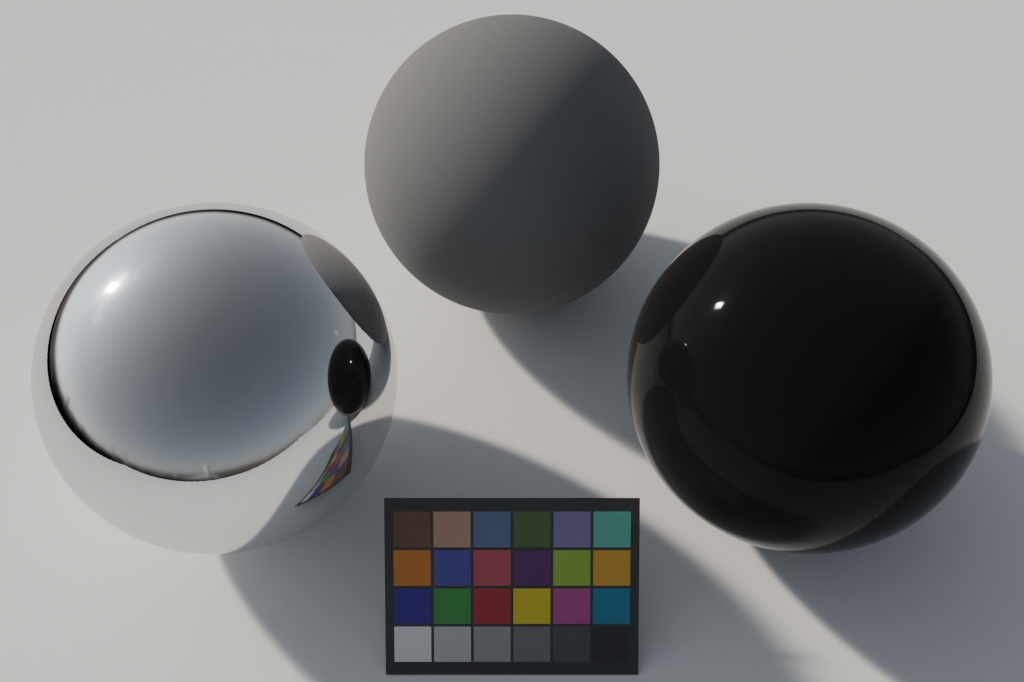} \\ 
    \rotatebox{90}{\parbox{1.5cm}{\centering {\textsf{mostly sunny}}}} & 
    \includegraphics[width=\tmplength]{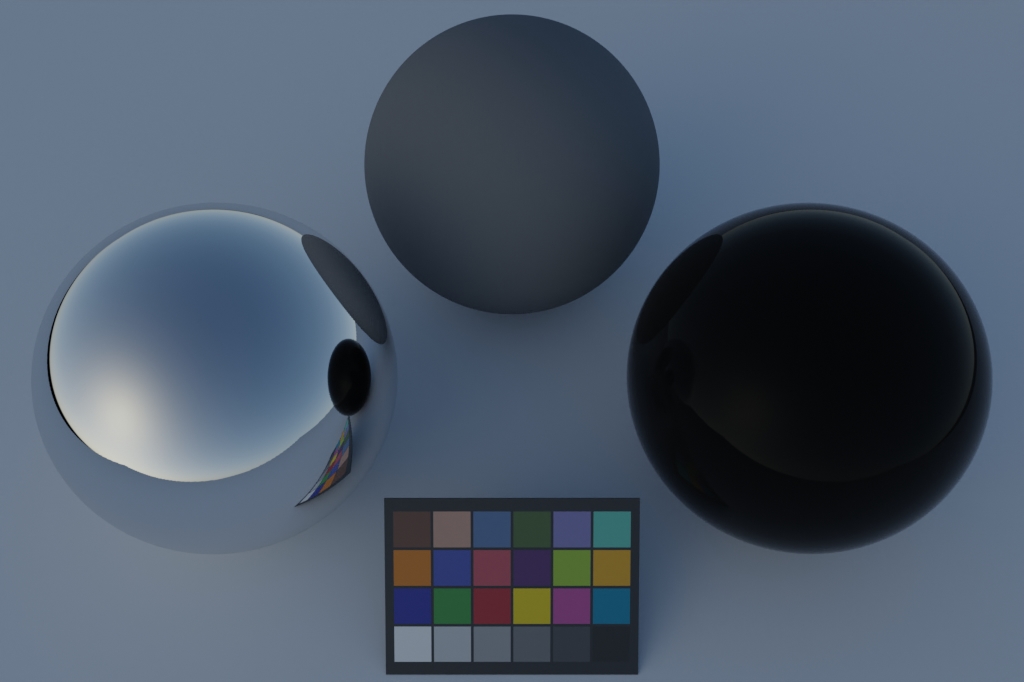} & 
    \includegraphics[width=\tmplength]{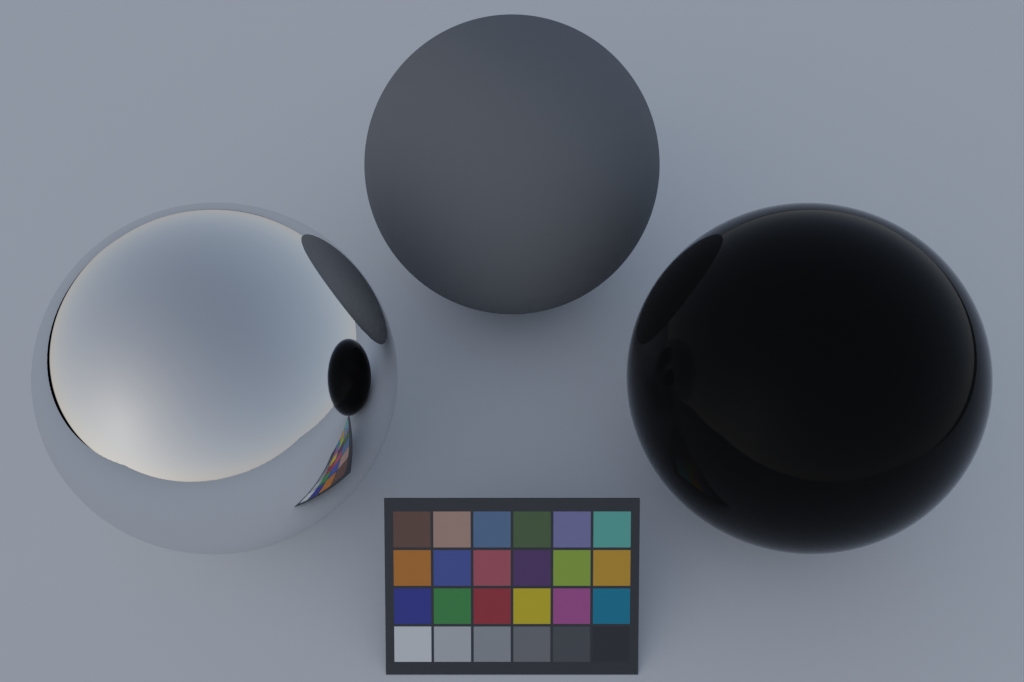} & 
    \includegraphics[width=\tmplength]{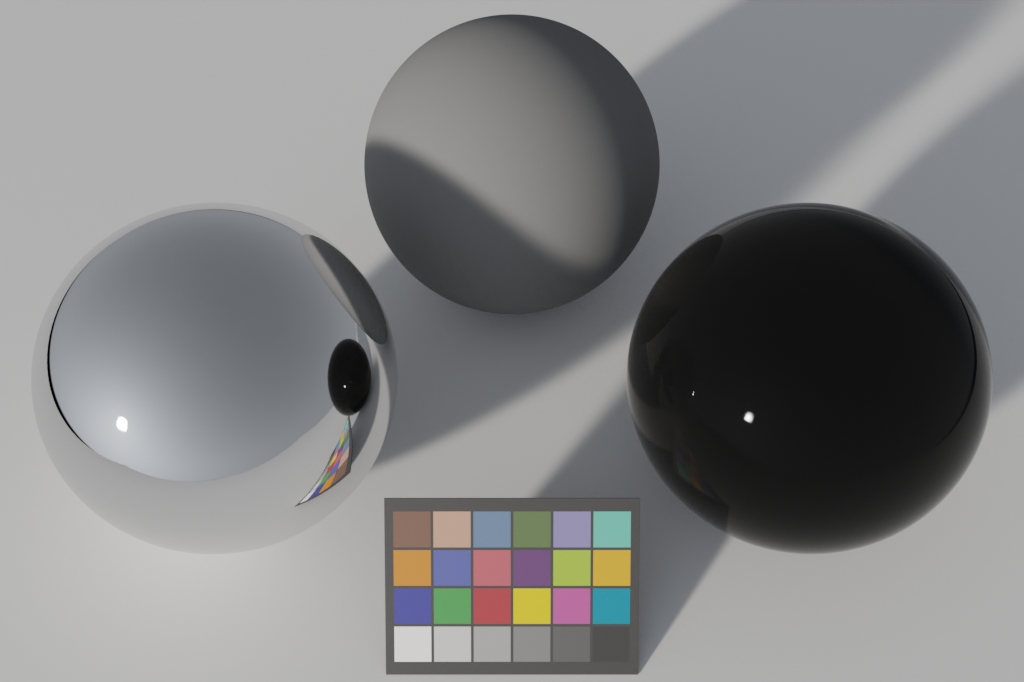} & 
    \includegraphics[width=\tmplength]{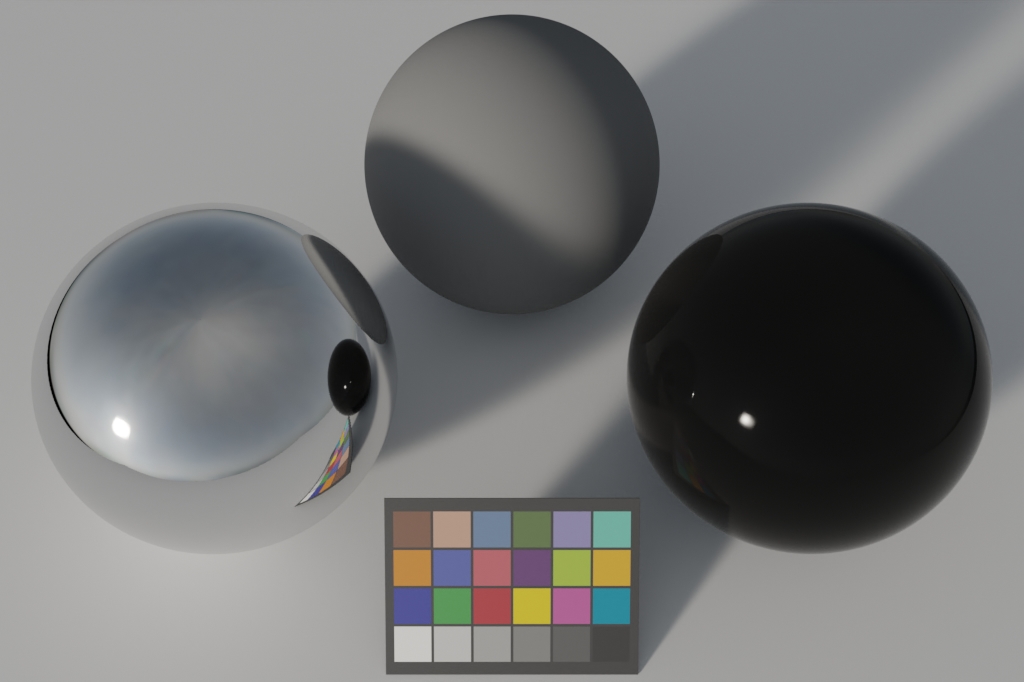} & 
    \includegraphics[width=\tmplength]{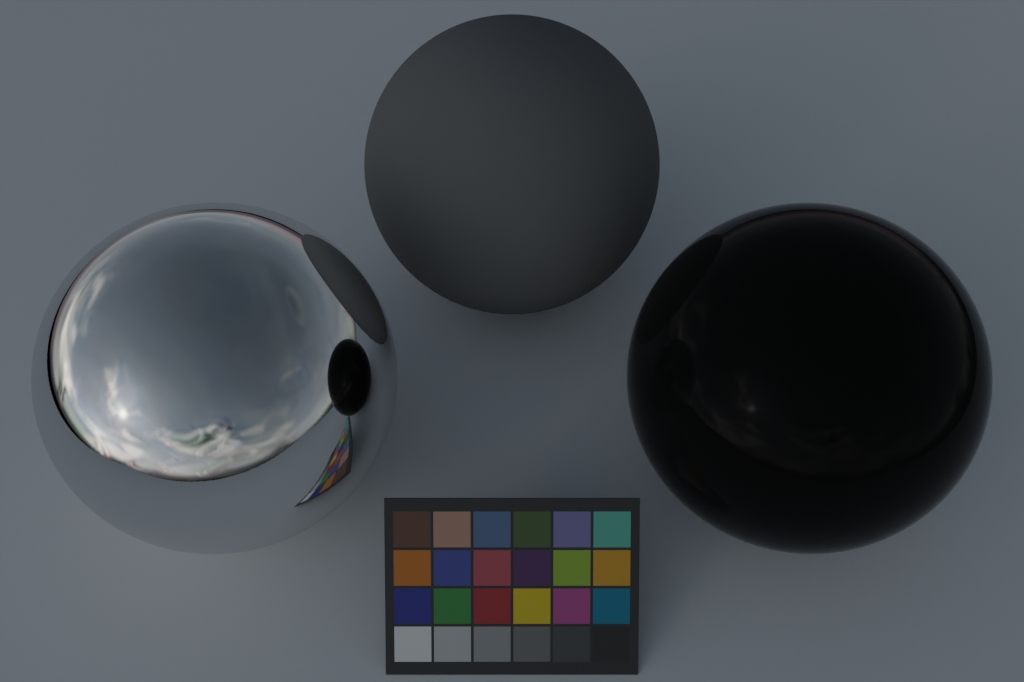} & 
    \includegraphics[width=\tmplength]{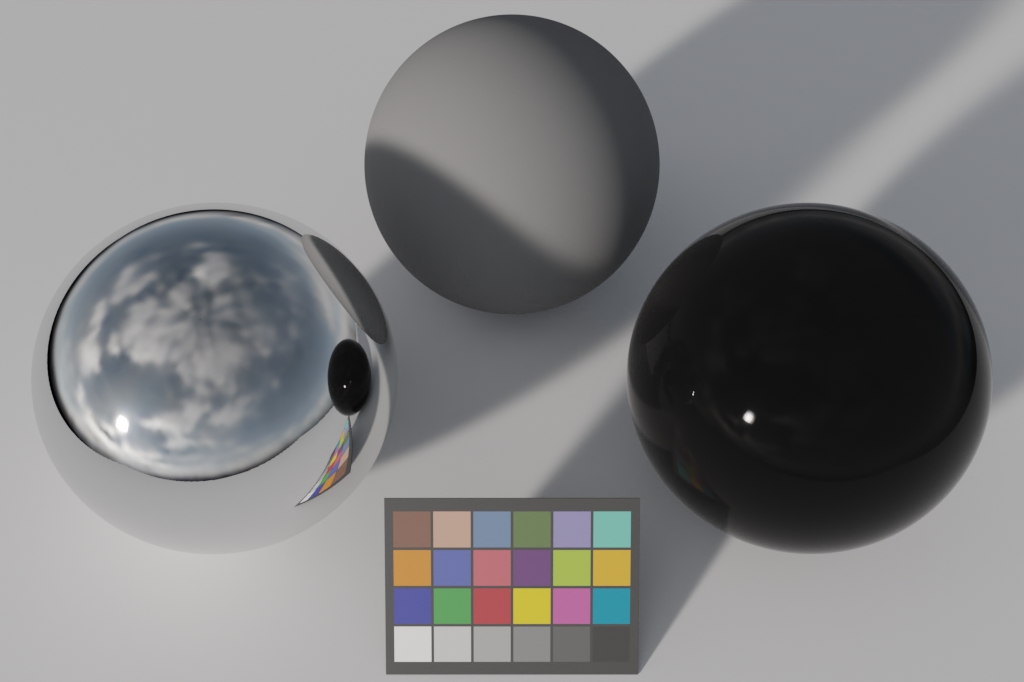} & 
    \includegraphics[width=\tmplength]{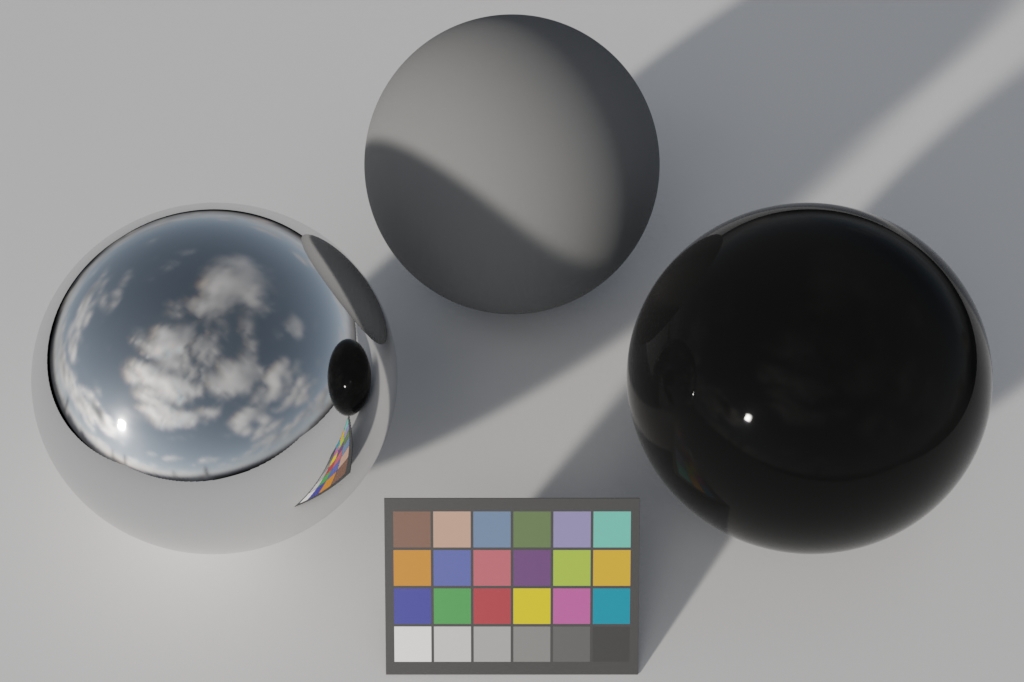} \\ 
    \rotatebox{90}{\parbox{1.5cm}{\centering {\textsf{partly cloudy}}}} & 
    \includegraphics[width=\tmplength]{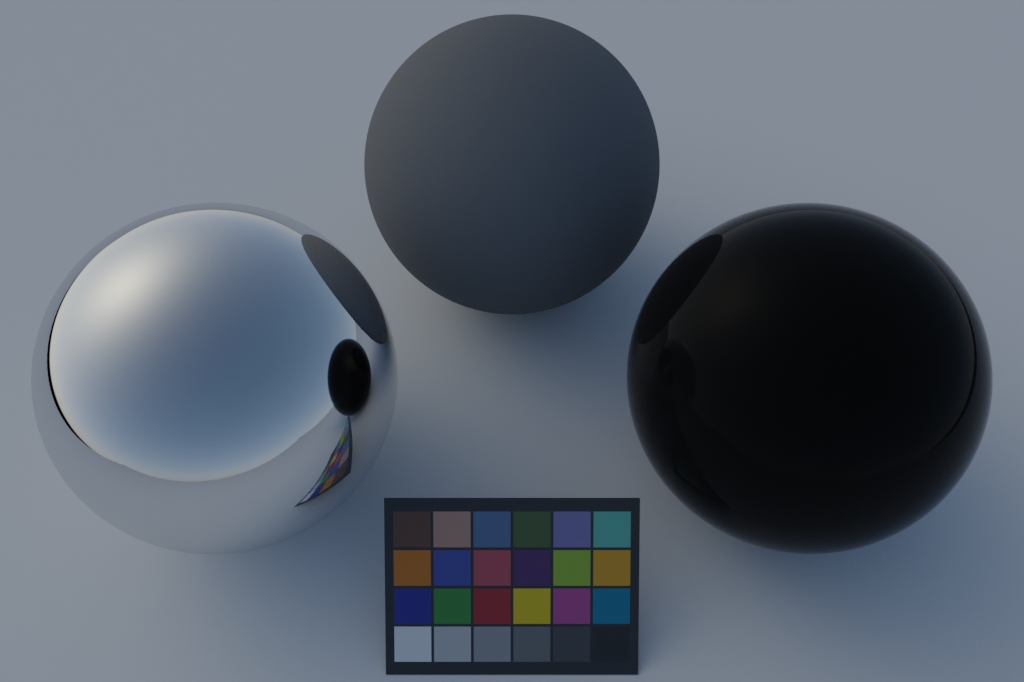} & 
    \includegraphics[width=\tmplength]{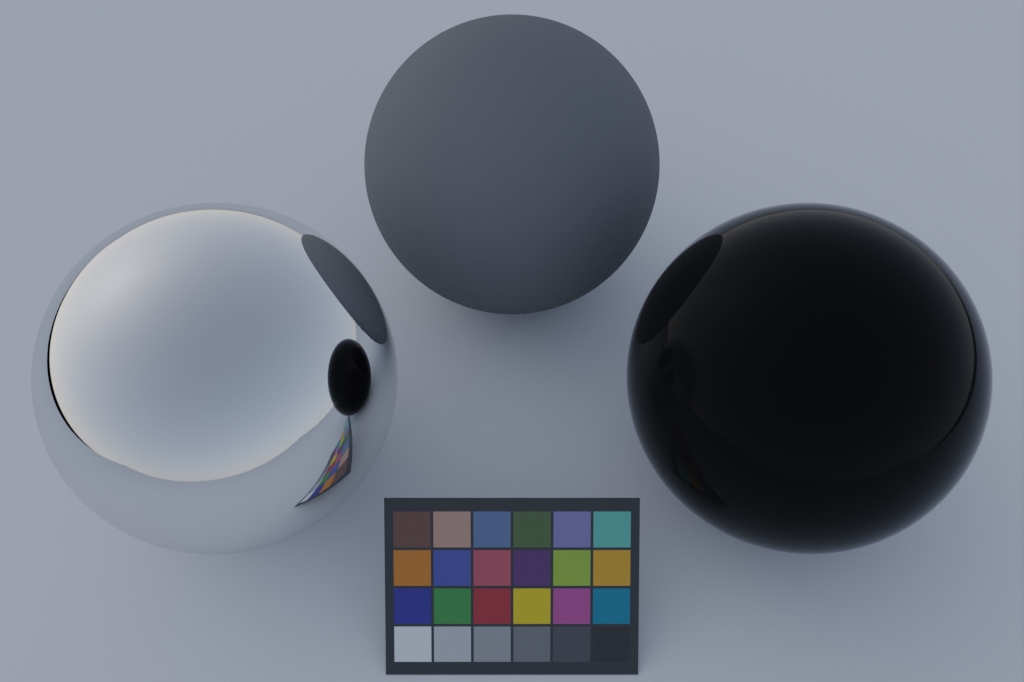} & 
    \includegraphics[width=\tmplength]{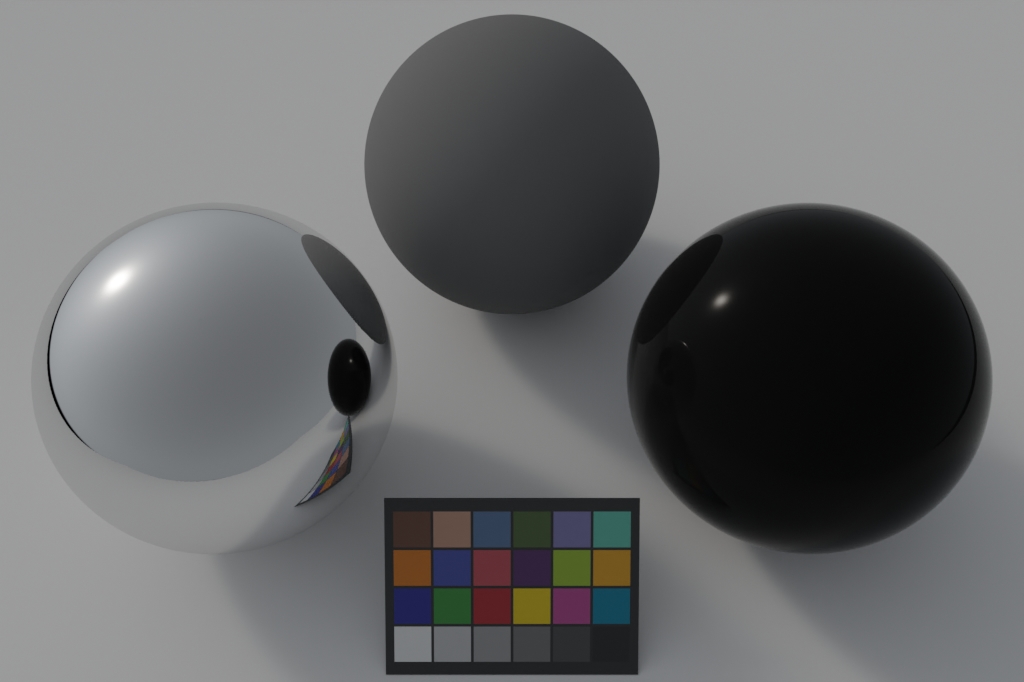} & 
    \includegraphics[width=\tmplength]{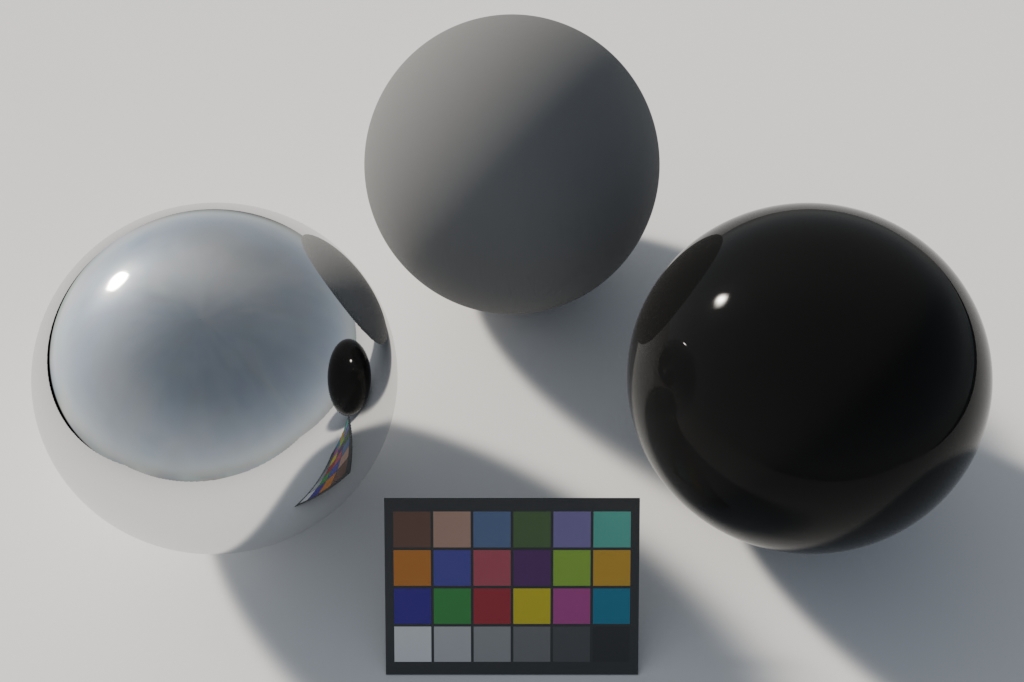} & 
    \includegraphics[width=\tmplength]{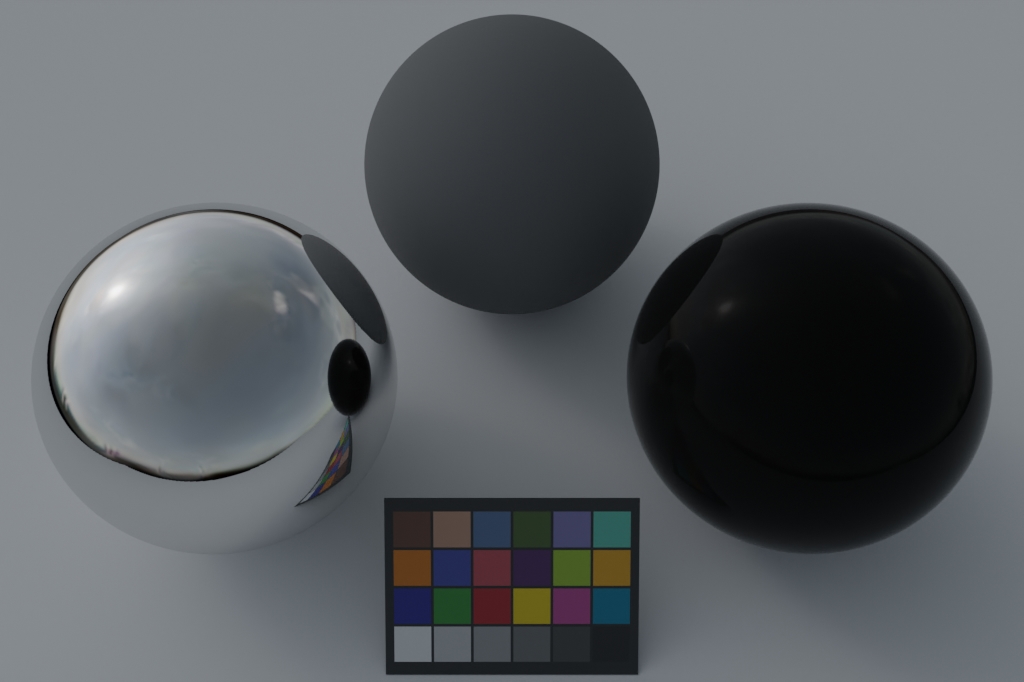} & 
    \includegraphics[width=\tmplength]{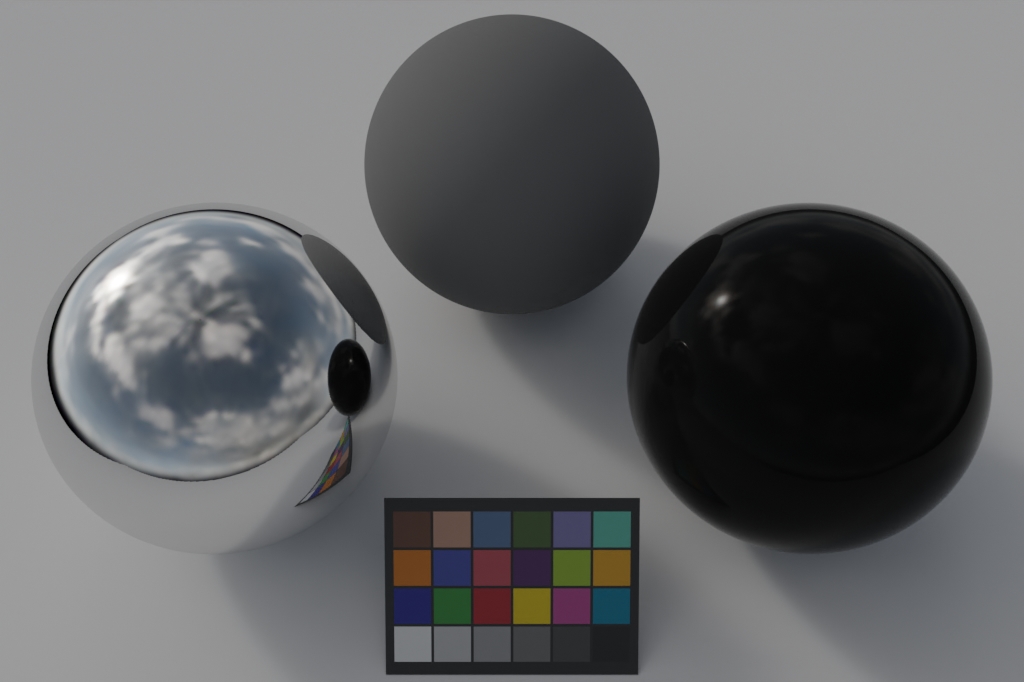} & 
    \includegraphics[width=\tmplength]{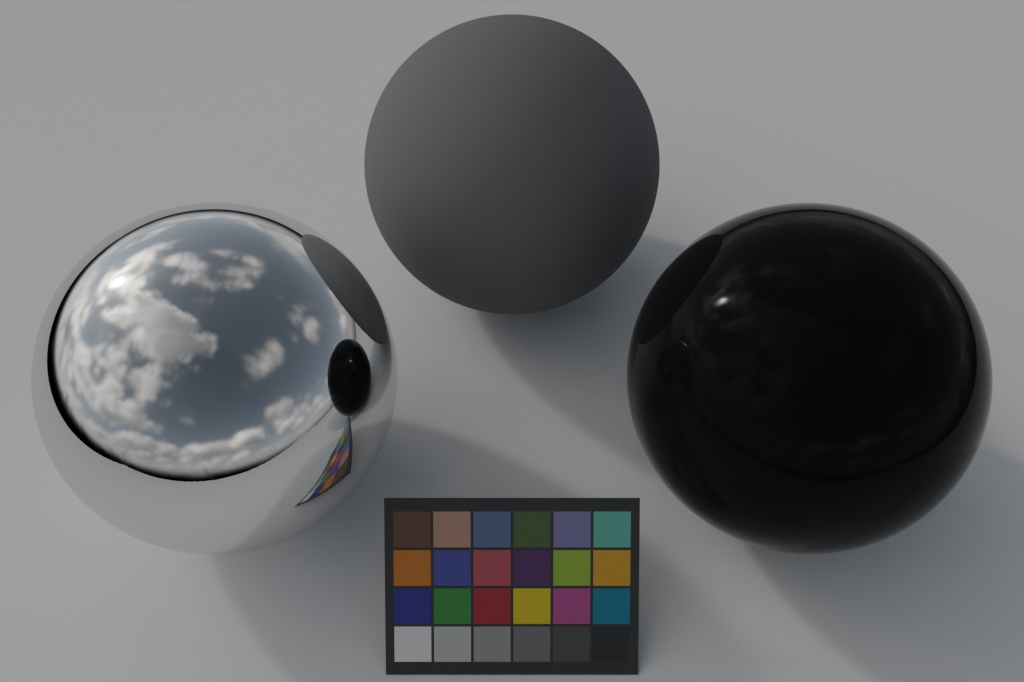} \\ 
    \rotatebox{90}{\parbox{1.5cm}{\centering {\textsf{mostly cloudy}}}} & 
    \includegraphics[width=\tmplength]{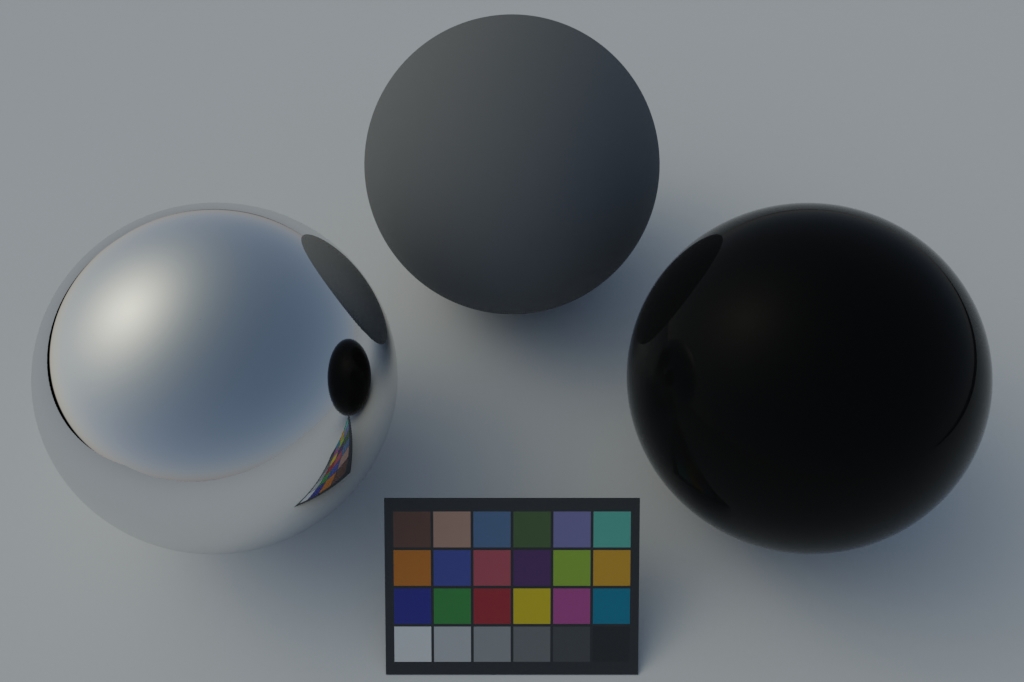} & 
    \includegraphics[width=\tmplength]{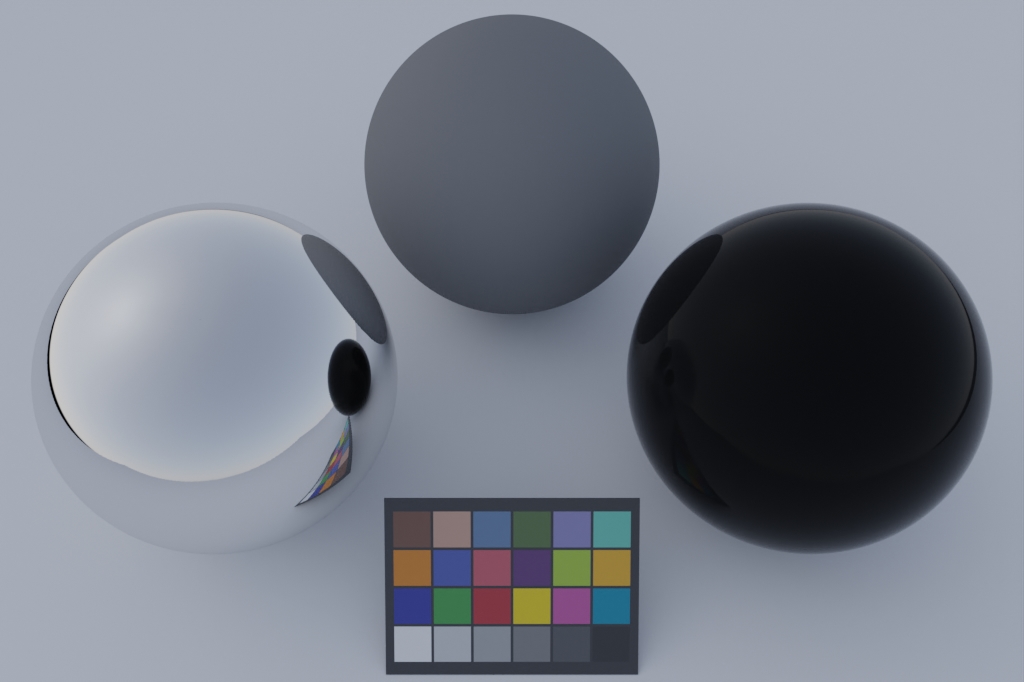} & 
    \includegraphics[width=\tmplength]{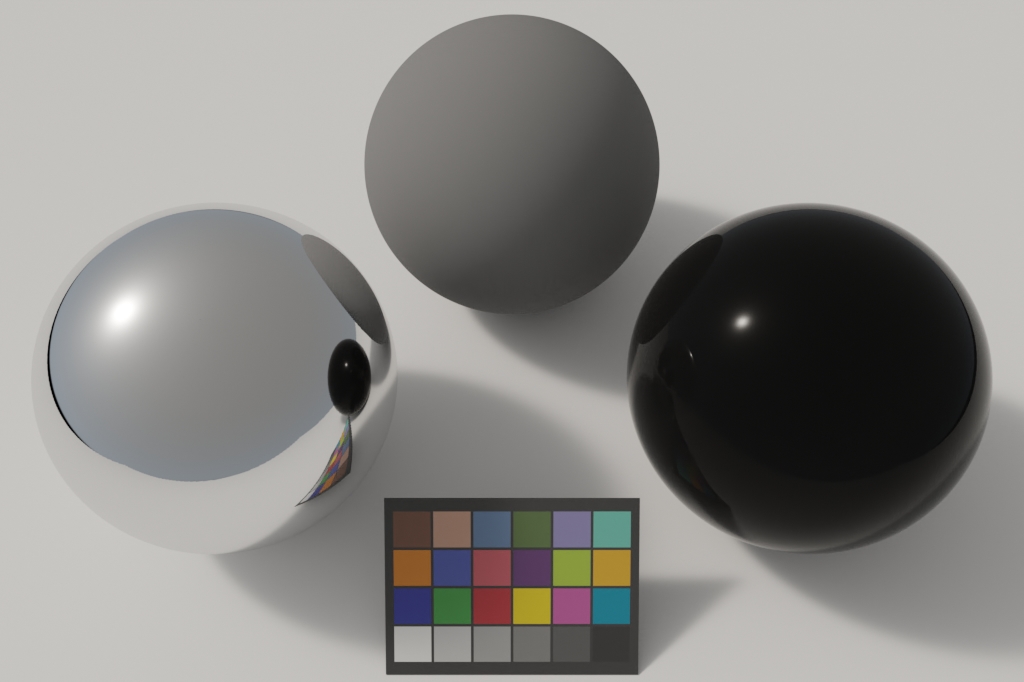} & 
    \includegraphics[width=\tmplength]{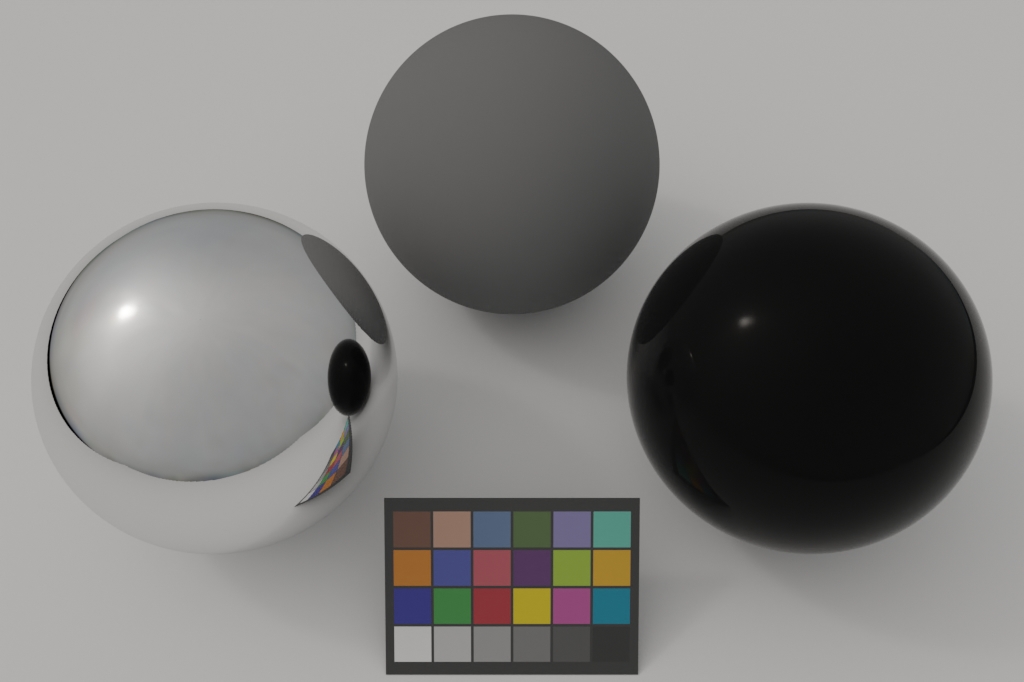} & 
    \includegraphics[width=\tmplength]{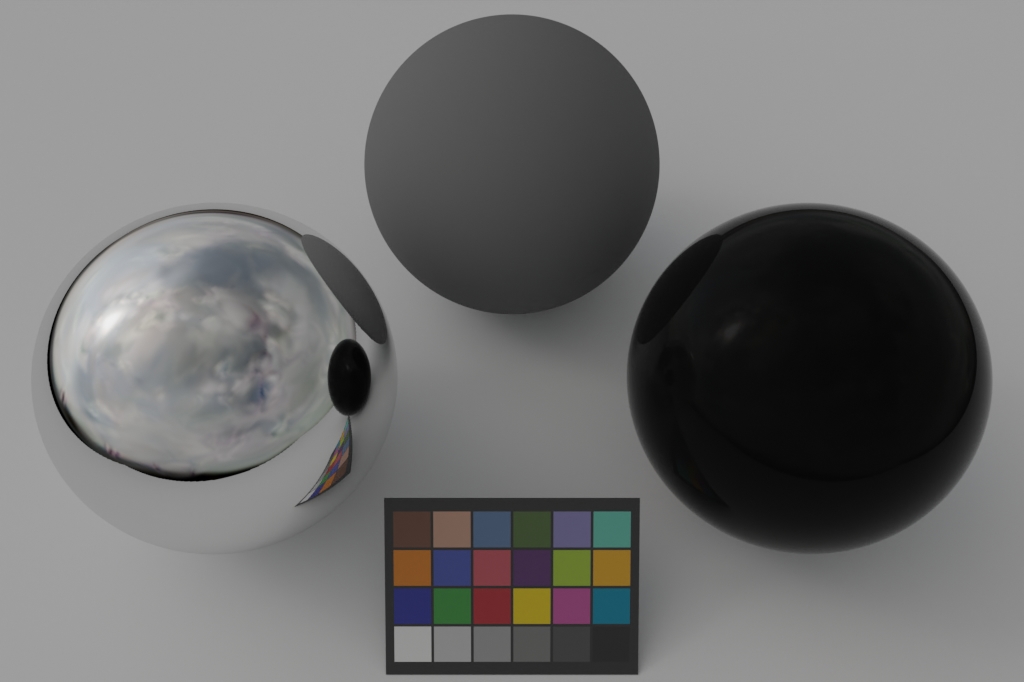} & 
    \includegraphics[width=\tmplength]{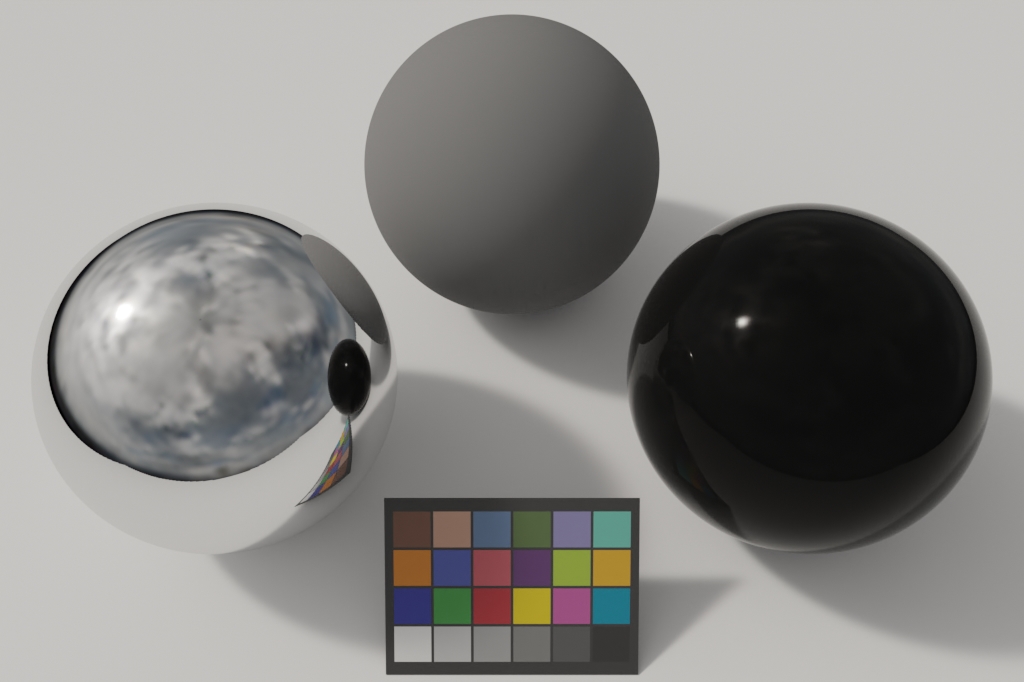} & 
    \includegraphics[width=\tmplength]{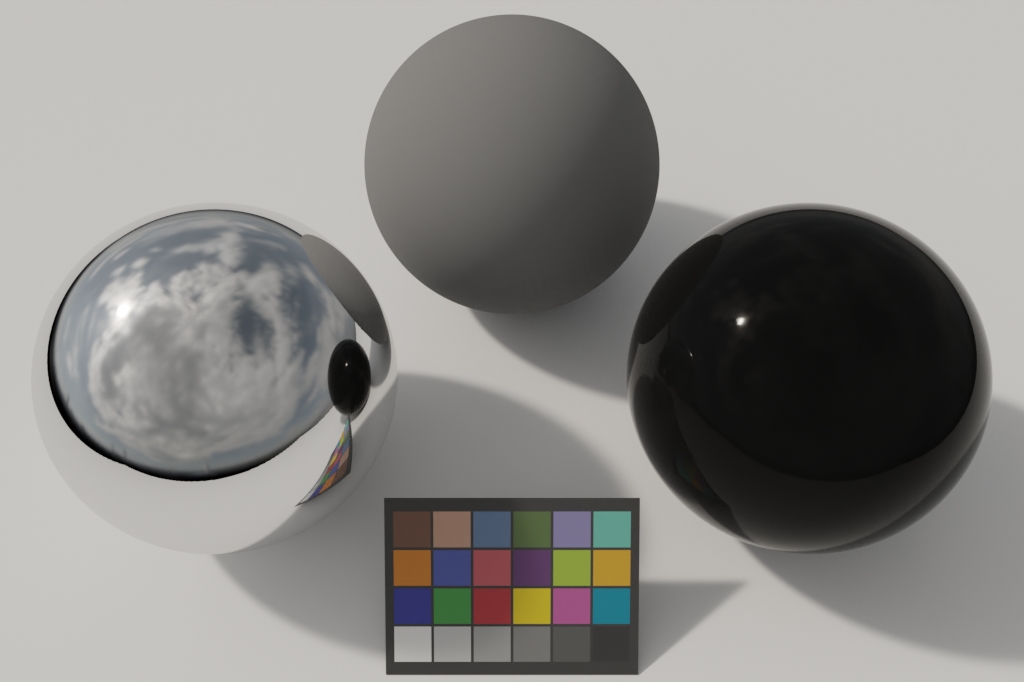} \\ 
    \rotatebox{90}{\parbox{1.5cm}{\centering {\textsf{overcast}}}} & 
    \includegraphics[width=\tmplength]{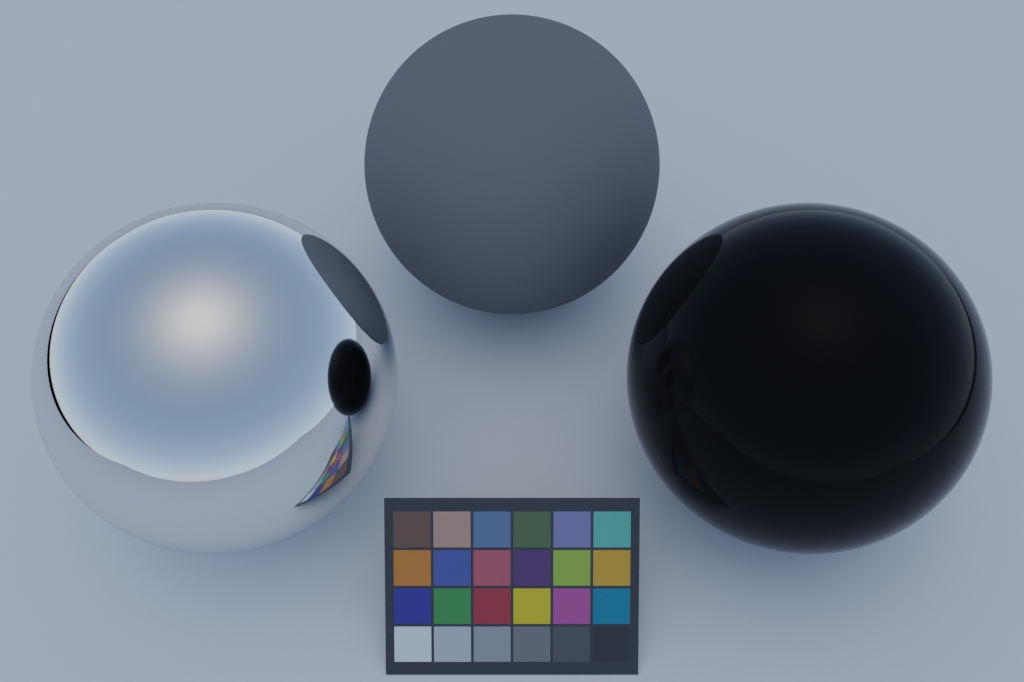} & 
    \includegraphics[width=\tmplength]{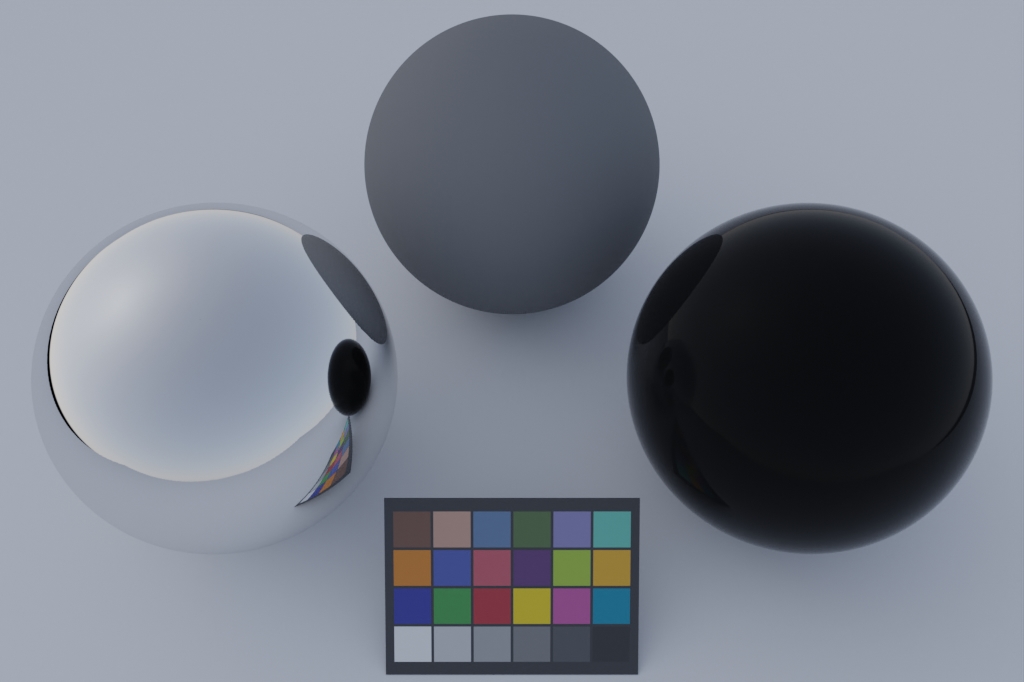} & 
    \includegraphics[width=\tmplength]{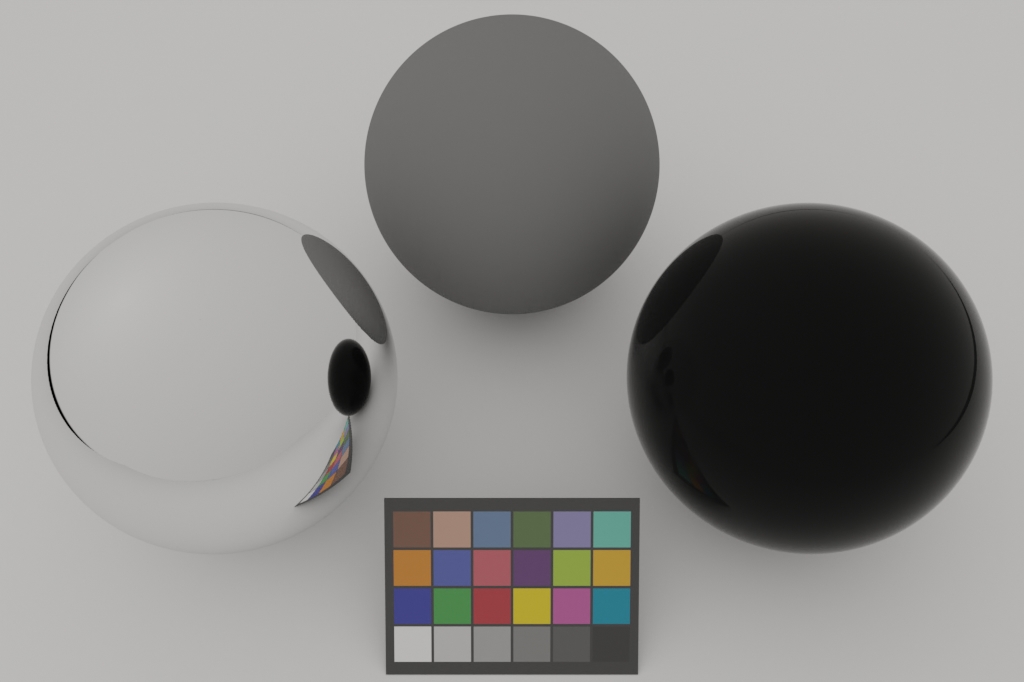} & 
    \includegraphics[width=\tmplength]{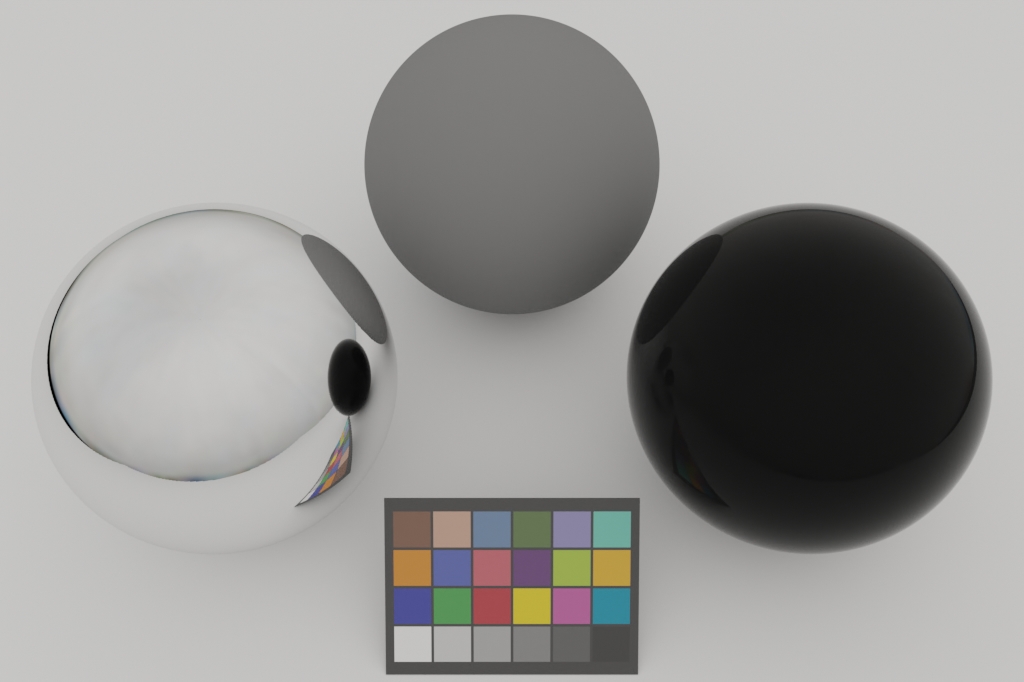} & 
    \includegraphics[width=\tmplength]{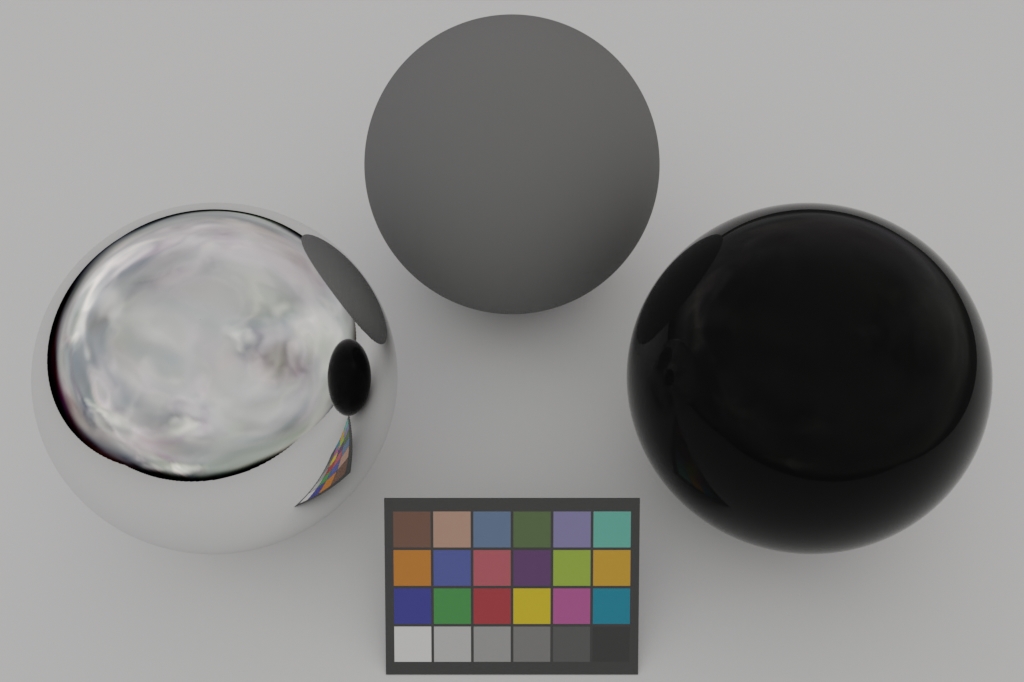} & 
    \includegraphics[width=\tmplength]{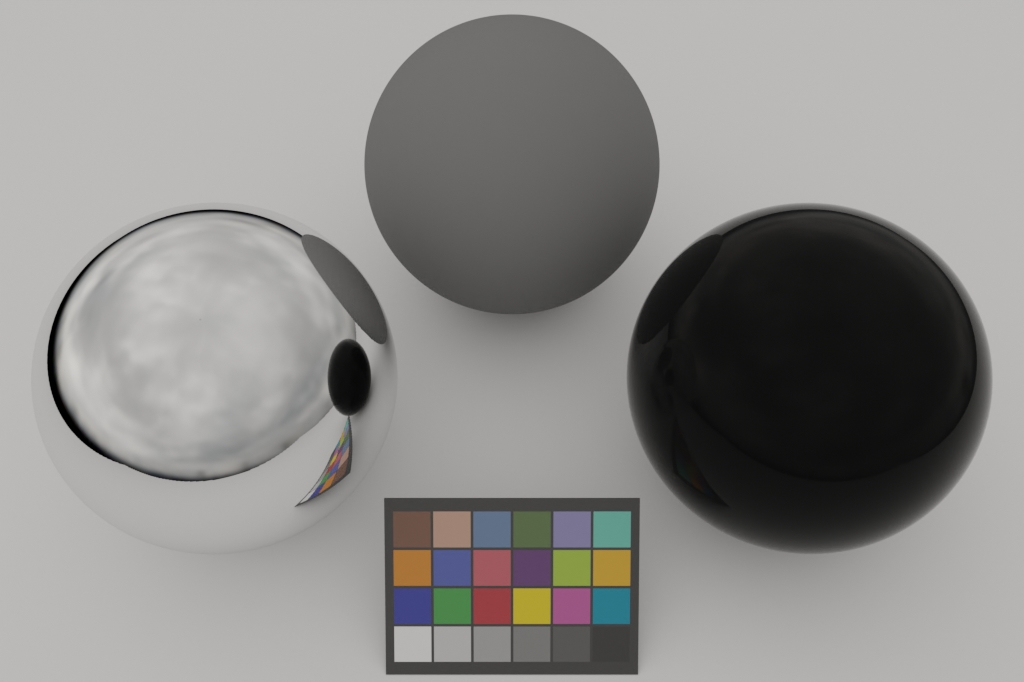} & 
    \includegraphics[width=\tmplength]{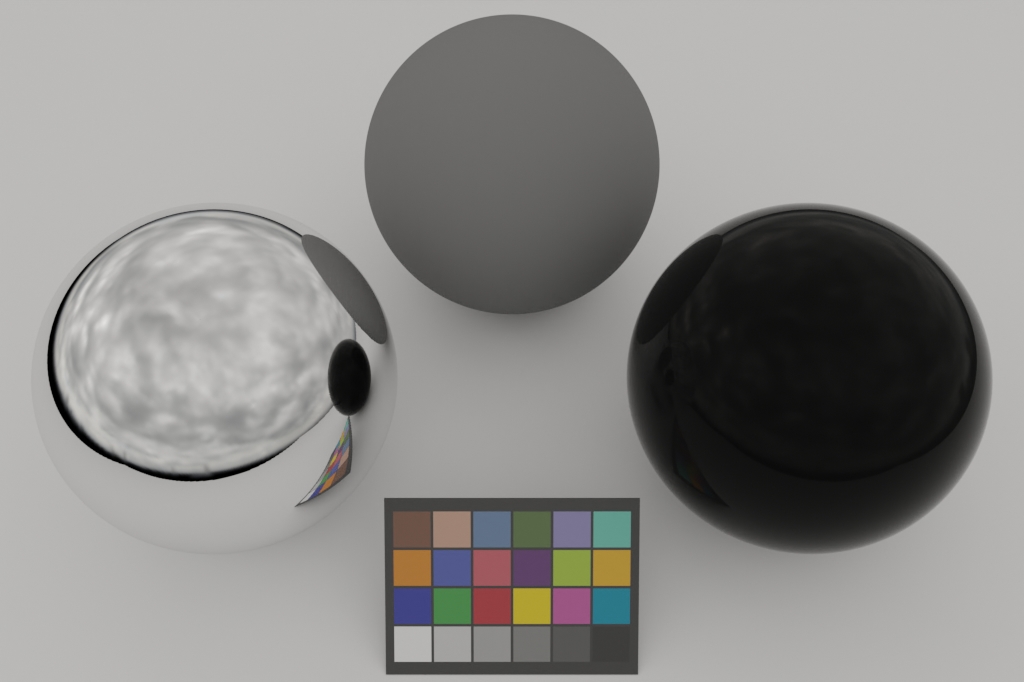} \\ 
    \end{tabular}
    \caption{\textbf{Visual comparison of a scene rendered with different lighting models.} Our approach generates high-quality reflections (shiny spheres) while preserving the proper energy (shading and shadows). The peak of HDR energy when the sun is visible can be verified in the black specular spheres.}
    \label{fig:renders-grid}
\end{figure*}

\subsubsection*{Qualitative evaluation}
We provide visual results from both parametric and learned sky models in \cref{fig:visuals-grid}. Our LM-GAN model yields the most plausible texture in terms of generated clouds for all levels of cloud coverage. While SkyGAN was also able to map different levels of cloud coverage, many artifacts are visible, which is coherent with the results shown in the original article. In this image, the advantage of using the LM model as input is also made clear, as its versatility enables it to emulate different levels of cloud coverage by changing the colors and sun falloffs of its clear skies, almost approximating SkyNet's auto-encoder results.
We further evaluate the dynamic range of the sky models with renders shown in \cref{fig:renders-grid}. In this experiment, we use the output of each method to light three spheres with different reflectance properties (mirror, Lambertian, glossy), each exhibiting some aspect of the environment: respectively, the texture, energy, and energy falloff around the sun region. In this case, the robustness of our LM fits becomes very clear, being visually enhanced by our model through textures, visible in the reflections.

\begin{figure}
    \centering
    \footnotesize
    \setlength{\tabcolsep}{.5pt}
    \renewcommand{\arraystretch}{0.4}
    \setlength{\tmplength}{0.19\linewidth}
    \begin{tabular}{ccccc}
        \textsf{{LM}~\shortcite{lm_model}} &
        \textsf{{LPIPS}} &
        \textsf{{LPIPS+GAN}} &
        \textsf{{full} (ours)} &
        \textsf{{reference}} \\
        \includegraphics[width=\tmplength]{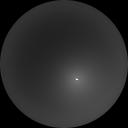} &
        \includegraphics[width=\tmplength]{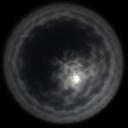} &
        \includegraphics[width=\tmplength]{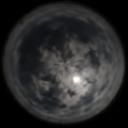} &
        \includegraphics[width=\tmplength]{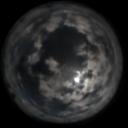} &
        \includegraphics[width=\tmplength]{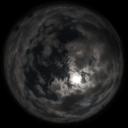} \\
        \includegraphics[width=\tmplength]{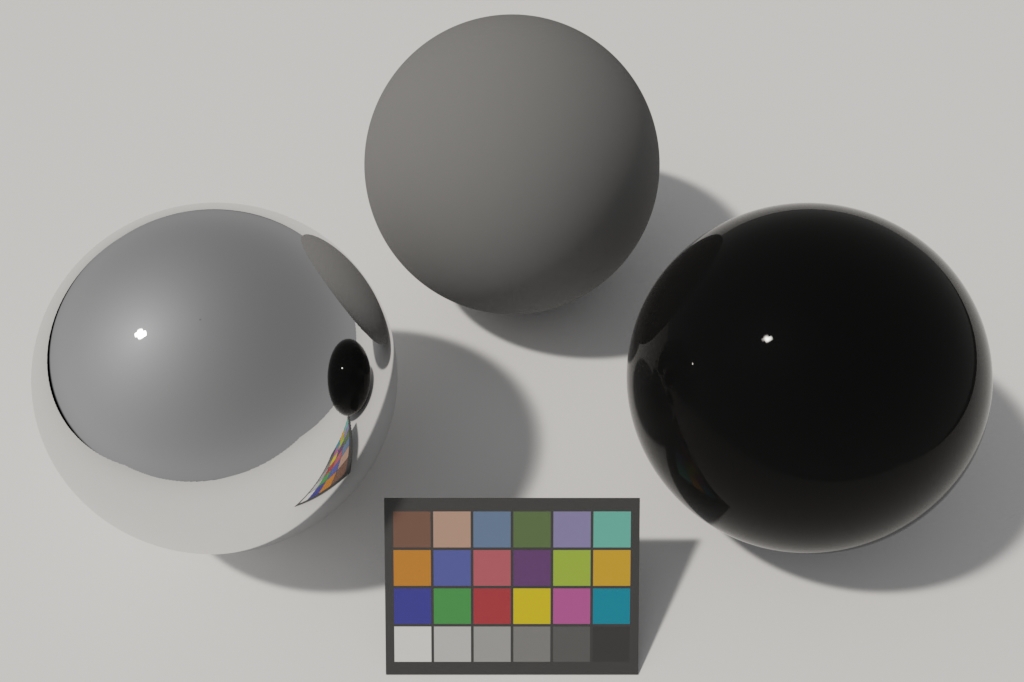} &
        \includegraphics[width=\tmplength]{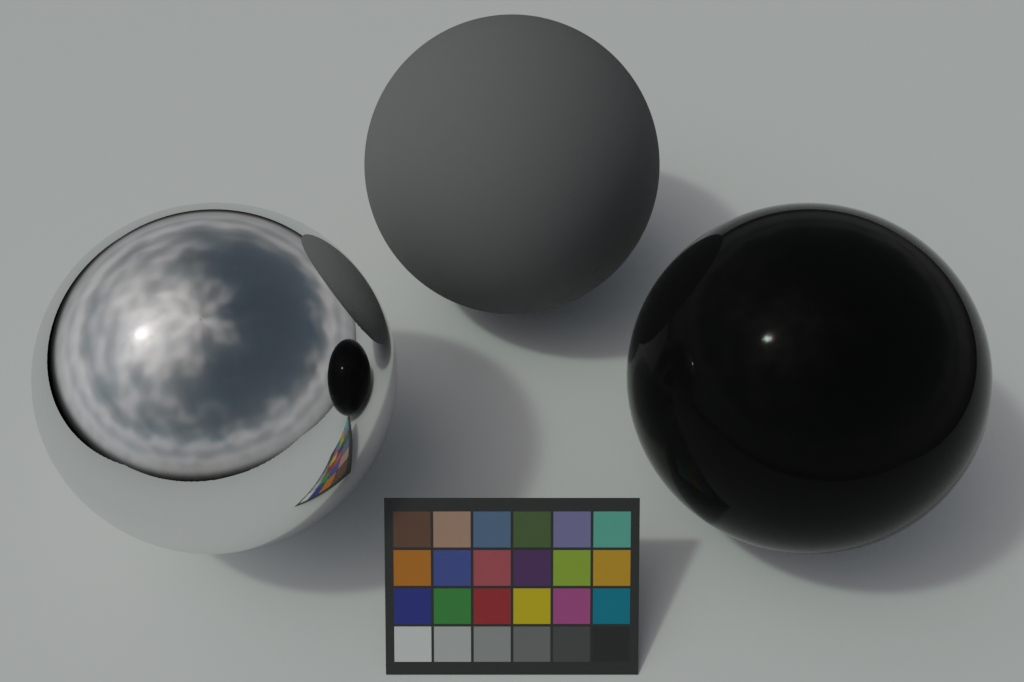} &
        \includegraphics[width=\tmplength]{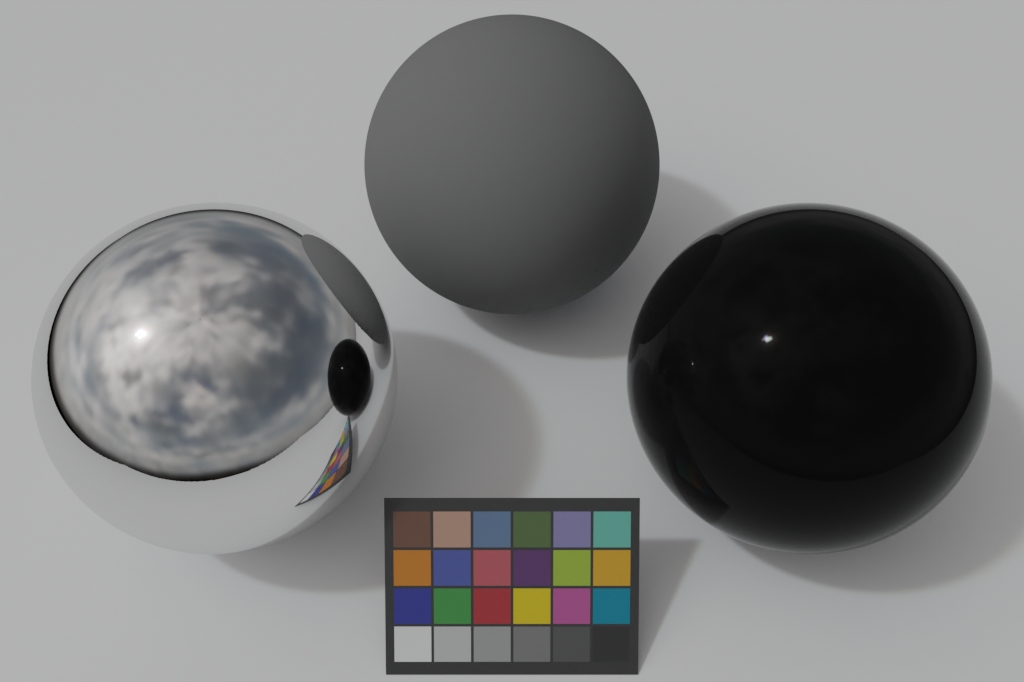} &
        \includegraphics[width=\tmplength]{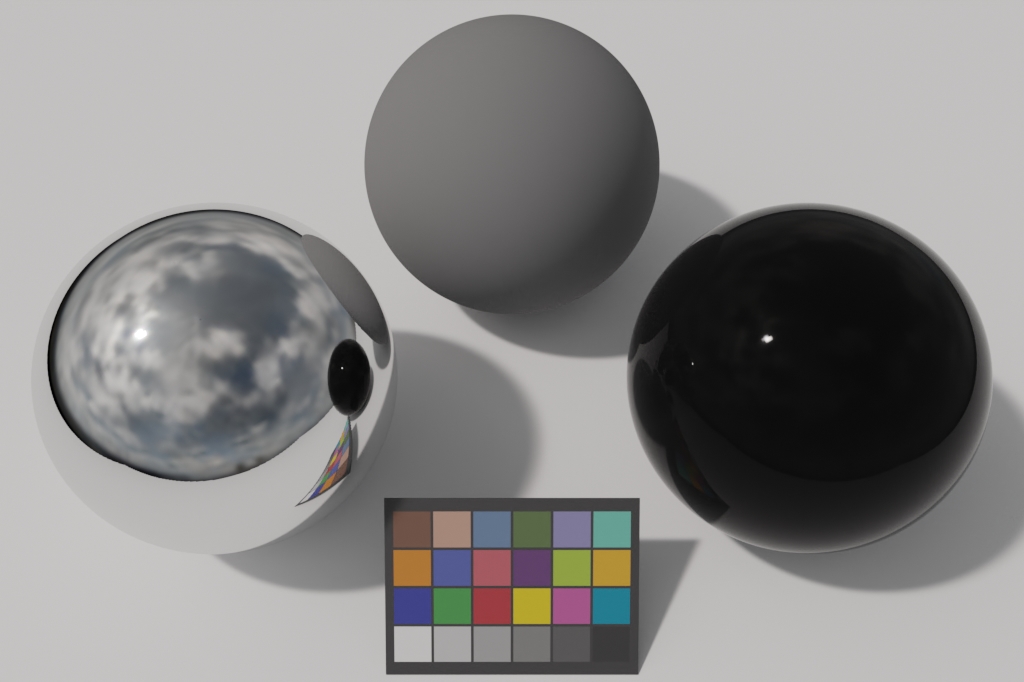} &
        \includegraphics[width=\tmplength]{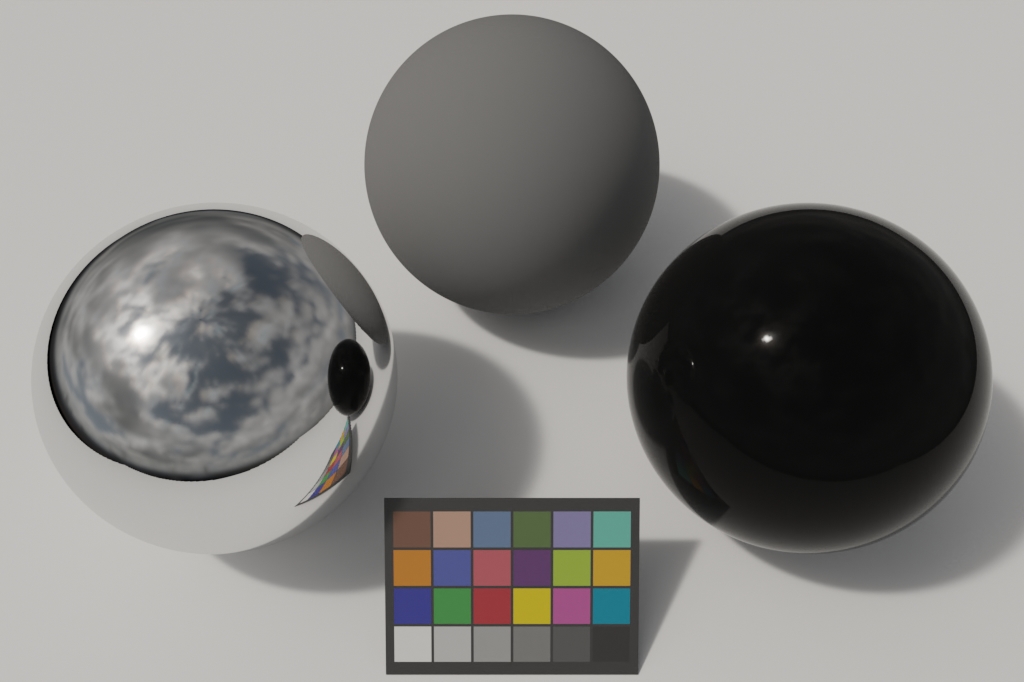} \\
    \end{tabular}    
    \caption{\textbf{Ablation example on a challenging sky.} Using an LPIPS loss (col. 2), the model learns high-frequency cloud textures, but not general structure. A patch discriminator (LPIPS+GAN) improves results (col. 3), but the upper dynamic range is still lost. Adding a rendering loss (full) (col. 4) helps correlate illumination (col. 1) and texture (col. 5).}
    \label{fig:ablation}
    \vspace{-4mm}
\end{figure}
\begin{table}[htbp]
    \setlength\tabcolsep{5pt}
    \centering

    \begin{tabular}{@{}l@{\hspace{.4\tabcolsep}}lccc@{}}
    & \small{\textsf{metric}}  & \small{\textsf{\textbf{LPIPS}}}   & \small{\textsf{\textbf{LPIPS+GAN}}}  & \small{\textsf{\textbf{full (ours)}}} \\
    \toprule
    \multirow{3}{*}{\rotatebox{90}{\footnotesize \textsf{texture}}} & 
    \textbf{\small{\textsf{FID}}}       & $67.09$    & $24.58$         & $\mathbf{20.28}$ \\
    & \textbf{\small{\textsf{RMSE}}}      & $11.11$              & $\mathbf{10.44}$ & $10.76$ \\
    & \textbf{\small{\textsf{si-RMSE}}}   & $10.18$              & $\mathbf{9.04}$    & $9.17$ \\
    \midrule
    \multirow{2}{*}{\rotatebox{90}{\footnotesize \textsf{ref.}}} & 
    \textbf{\small{\textsf{RMSE}}}       & $1275.40$ & $1094.30$ & $\mathbf{609.94}$ \\
    & \textbf{\small{\textsf{si-RMSE}}}   & $602.24$ & $261.41$ & $\mathbf{122.06}$ \\
    \midrule
    \multirow{2}{*}{\rotatebox{90}{\footnotesize \textsf{input}}} & 
    \textbf{\small{\textsf{RMSE}}}     & $1215.82$ & $911.99$ & $\mathbf{142.43}$ \\
    & \textbf{\small{\textsf{si-RMSE}}}   & $617.76$ & $261.34$ & $\mathbf{69.65}$ \\
    \bottomrule
    \end{tabular}
    \caption{\textbf{Ablation errors.} Adding a GAN loss to an LPIPS-based UNet significantly improved visual quality. There was also a significant gain in scale-invariant lighting accuracy, showing the influence texture has on contrast and patterns. Adding a rendering loss helped modulate the remaining lighting loss while improving visual fidelity. }
    \label{tab:ablation}
    \vspace{-5mm}
\end{table}

\subsection{Ablation studies}
\label{sec:ablations}

To identify the contribution of the individual losses to the final result, we train two other models for the same amount of epochs: one solely on an LPIPS loss \cite{lpips}, the other with LPIPS and the adversarial PatchGAN loss (without the rendering loss). All models were trained at $128 \times 128$ resolution. \Cref{fig:ablation} shows the LPIPS model captures cloud textures locally, repeating it over the image like a pattern, however fails to provide both photorealistic and HDR results. Combining it with PatchGAN is sufficient to achieve a good appearance, but it still struggles to preserve the full HDR, improving only scale-invariant results through proper texture (see tab.~\cref{tab:ablation}). Adding a rendering loss further improved the texture and the desired lighting properties, enforcing the desired correlation.
\section{Conclusion}

We present LM-GAN, a learned HDR sky model that generates photorealistic sky environment maps, readily usable for lighting virtual scenes and objects. Our key contribution lies in the learning of both energy and texture from the sky and the sun simultaneously into a single standalone model, trained directly on real captured skies. 
Our method is implemented via a rejuvenated yet simple UNet architecture, which is straightforward to implement and efficient to execute. Our proposed model is conditioned on a parametric sky model, which confers both expressivity and physical accuracy to our results. 
This results in a flexible and user-editable sky model that can be used for several applications. 

Despite our state-of-the-art results, our method bears some limitations.
First, our method cannot provide the same resolution as professionally captured HDRIs (commonly 12k x 24k) due to the limitations of generative machine learning algorithms and the currently available hardware. 
Also, while our model has several user-editable characteristics such as the sun position and intensity and the color of the sky, our proposed method does not allow for fine-grained control of cloud appearance and location. 

We hope our work paves the way for high-quality end-to-end environment modeling which captures the richness of the natural sky, eventually replacing manual artist annotations when lighting 3D scenes or editing images in a photorealistic way.

\clearpage
\section*{Acknowledgements}

This work was supported by NSERC grants ALLRP 557208-20, RGPIN-2020-04799, and Adobe. The authors also thank Nvidia for donating GPUs; Tobias Rittig and Martin Mirbauer for running their Prague sky model fitting code on our dataset; Jinsong Zhang, Henrique Weber, and Aryan Garg for their help with various parts of the code; and Heitor Felix for helpful discussions.

\include{sections/6-page_figures.tex}

\end{document}